\documentclass[reprint,secnumarabic,nobibnotes,aps,prb,superscriptaddress]{revtex4-2}

\usepackage{color, xcolor, colortbl}
\usepackage{graphicx}
\usepackage{amsthm,amsmath,amssymb,amsfonts}
\usepackage{algorithm}
\usepackage{algorithmic}
\usepackage{dcolumn}
\usepackage{epstopdf}
\usepackage{bm}
\usepackage[caption=false]{subfig}
\usepackage{appendix}
\usepackage{multirow}
\usepackage{braket}
\usepackage[english]{babel}
\usepackage{hyperref}
\usepackage[capitalize]{cleveref}
\usepackage{xpatch}
\usepackage{tikz}
\usepackage{adjustbox}
\usepackage{xspace}

\renewcommand{\Im}{\operatorname{Im}}

\newcommand{\abs}[1]{\left\lvert#1\right\rvert}
\newcommand{\norm}[1]{\left\lVert#1\right\rVert}

\newcommand{\Or}{\mathcal{O}}

\providecommand{\definitionname}{Definition}
\providecommand{\assumptionname}{Assumption}
\providecommand{\corollaryname}{Corollary}
\providecommand{\lemmaname}{Lemma}
\providecommand{\propositionname}{Proposition}
\providecommand{\remarkname}{Remark}
\providecommand{\theoremname}{Theorem}

\usetikzlibrary{fit}
\tikzset{%
  highlight/.style={rectangle,rounded corners,fill=blue!15,draw,fill opacity=0.3,thick,inner sep=0pt}
}

\usepackage{pifont}
\newcommand{\cmark}{{\color{green}\ding{51}}}%
\newcommand{\xmark}{{\color{red}\ding{55}}}%

\newcommand{\DeptMath}{Department of Mathematics, University of California, Berkeley, California 94720 USA}
\newcommand{\LBLMath}{Applied Mathematics and Computational Research Division, Lawrence Berkeley National Laboratory, Berkeley, CA 94720, USA}

\begin{document}

\title{Robust analytic continuation of Green's functions via projection, pole estimation, and semidefinite relaxation}
\author{Zhen Huang}
\affiliation{\DeptMath}
\author{Emanuel Gull}
\affiliation{Department of Physics, University of Michigan, Ann Arbor, Michigan 48109, USA}
\author{Lin Lin}
\email{linlin@math.berkeley.edu}
\affiliation{\DeptMath}
\affiliation{\LBLMath}
\date{\today}

\begin{abstract}
Green's functions of fermions are described by matrix-valued Herglotz-Nevanlinna functions.
Since analytic continuation is fundamentally an ill-posed problem, the causal space described by the matrix-valued Herglotz-Nevanlinna structure can be instrumental in improving the accuracy and in enhancing the robustness with respect to noise. We demonstrate a three-pronged procedure for robust analytic continuation called PES: (1) Projection of data to the causal space. (2) Estimation of pole locations. (3) Semidefinite relaxation within the causal space. We compare the performance of PES with the recently developed Nevanlinna and Carath\'{e}odory  continuation methods and find that PES is more robust in the presence of noise and does not require the usage of extended precision arithmetics. We also demonstrate that a causal projection improves the performance of the Nevanlinna and Carath\'{e}odory methods.
The PES method is generalized to bosonic response functions, for which the Nevanlinna and Carath\'{e}odory  continuation methods have not yet been developed.
It is particularly useful for studying spectra with sharp features, as they occur in the study of molecules and band structures in solids.
\end{abstract}

\maketitle

\section{Introduction}
Green's functions of  quantum many-body systems are central subjects in condensed matter physics, quantum chemistry, and quantum field theory. Due to the difficulty of performing correlated 
finite-temperature calculations on the real axis, numerical methods typically obtain results on the Matsubara axis and employ analytic continuation as a postprocessing tool to infer Green's functions along the real axis. 
Analytic continuation is fundamentally an ill-posed problem, and many numerical methods have been developed, such as the Pad\'{e} fit \cite{BakerBakerJrGraves-MorrisEtAl1996,SchottLochtLundin2016},
 maximum entropy (MaxEnt) \cite{JarrellGubernatis1996,AsakawaNakaharaHatsuda2001,LevyLeBlancGull2017,Bryan1990,Rothkopf2020},  stochastic analytic continuation  and its variants \cite{Sandvik1998,VafayiGunnarsson2007,FuchsPruschkeJarrell2010,ShaoSandvik2022},  sparse modeling \cite{YoshimiOtsukiMotoyamaEtAl2019,MotoyamaYoshimiOtsuki2022}, and machine
learning approaches \cite{FournierWangYazyevEtAl2020,YoonSimHan2018}.
All these methods encounter difficulties in satisfying causality, recovering sharp features, resolving multiple features, and/or capturing high frequency information in the spectra. 
Furthermore, most methods have been developed only for diagonal entries of the Green's function. The analytic continuation of  off-diagonal entries of the Green's function \cite{TomczakBiermann2007,DangAiMillisetal2014,GullMillis2014,krabergerTrieblZingletal2017,ReymbautTremblay2017} is often attempted by diagonalizing the Green’s function at a certain Matsubara frequency, conducting the analytic continuation only of the diagonal entries with respect to the transformed basis at other frequencies, and neglecting the remaining non-diagonal entries.

The fermionic Green's functions is a special type of  \textit{matrix-valued Herglotz-Nevanlinna functions}. These functions have a wide range of  scientific and engineering applications, and have been (somewhat inconsistently) named after renowned mathematicians including Carath\'{e}odory, Herglotz, Nevanlinna, Pick, and Riesz. They are also sometimes called $R$-functions \cite{KacKrein1974}. The name  matrix-valued Herglotz-Nevanlinna functions adopted in this paper follows the suggestion in \cite{BelyiHassiSnooEtAl2006}.
This crucial analytic structure has only been taken into account recently by the Nevanlinna continuation method \cite{FeiYehGull2021} for diagonal entries of Green's functions, and by the Carath\'{e}odory method \cite{FeiYehZgidEtAl2021} for both diagonal and off-diagonal entries.
In the absence of noise, these methods have reached unprecedented accuracy in analytic continuation by resolving complicated spectral functions with multiple features.
However, these methods also have notable drawbacks. (1) They are not numerically stable, and extended precision arithmetic operations (128 bits of precision or higher) are needed to carry out such calculations (even in the absence of noise). Therefore the computational cost of the analytic continuation (especially for the Carath\'{e}odory continuation) can be large. (2) The Nevanlinna/Carath\'{e}odory interpolants exist if and only if  Pick's criterion \cite{Pick1917,KhargonekarTannenbaum1985,FeiYehGull2021} is satisfied by the Matsubara data. In practice, noisy data  often violates  Pick's
criterion, meaning that the computational result is not guaranteed to be causal. (3) In their standard form,  matrix-valued Herglotz-Nevanlinna structure is only applicable to fermions, and hence these methods are not directly applicable to bosonic systems. 

The Herglotz-Nevanlinna function and its interpolation have been thoroughly studied in areas such as the control theory (see e.g., \cite{Tannenbaum1987} and \cite[Chapter 9.2]{DoyleFrancisTannenbaum2013}). 
The problem of analytic continuation is also intimately related to problems in signal processing, and in particular fitting with exponential functions  (see e.g., \cite{OsborneSmyth1995,GolubPereyra2003,BeylkinMonzon2005}). 
One advantage of Nevanlinna and Carath\'{e}odory methods is that they require little prior knowledge of the structure of Green's functions.
With further prior knowledge of the lower and upper bound of the absolute value of the pole locations, Ying recently showed that a modified Prony's method  can be used for performing analytic continuation with noisy data with multiple features using only double precision arithmetic operations~\cite{Ying2022analytic,Ying2022pole}.

In this article, we propose a three-pronged Projection-Estimation-Semidefinite relaxation (PES) method to perform analytic continuation within the causal space, while avoiding the aforementioned drawbacks.
The meaning of the three steps are as follows: (1) Projection of the noisy data into the causal space; (2) Estimation of pole locations using the adaptive Antoulas--Anderson (AAA) algorithm~\cite{NakatsukasaSeteTrefethen2018}; (3) Semidefinite relaxation (SDR) fitting of Green's functions (diagonal and off-diagonal elements) using a bi-level optimization approach \cite{Mejuto-ZaeraZepeda-NunezLindseyEtAl2020}. Each step of the approach aims at resolving certain aspects of the difficulties in the analytic continuation of noisy Matsubara Green's functions.
The prior knowledge needed for the PES method is comparable to that in Nevanlinna and Carath\'{e}odory methods.
We demonstrate that the PES method can robustly perform analytical continuation of noisy data using standard double precision arithmetic operations, and is applicable to both fermionic and bosonic systems. 
Table~\ref{table:comparison} compares the PES method with the Nevanlinna, Carath\'{e}odory,  MaxEnt, and Pad\'{e} method.  
We  emphasize that each of the three steps of PES can also be useful in improving the robustness of other analytic continuation methods. For instance, the projection step can be employed as a data pre-processing step that significantly  improves the robustness of the Nevanlinna and Carath\'{e}odory continuation for noisy data. The semidefinite relaxation can be combined with  other continuation methods as a post-processing refinement.

\begin{table*}[ht]
\begin{tabular}{|c|c|c|c|c|c|}
\hline
\hline
Method       & 
$\begin{array}{c}
     \text{Noise} \\
     \text{Robustness} 
\end{array}$&
${\begin{array}{c}
      \text{Calculation}\\
      \text{ Precision }\\
      \text{Requirement}
\end{array}}$ 
& $\begin{array}{c}
     \text{Sharp} \\
     \text{Features} 
\end{array}$ 
&Causality & Applicability\\ \hline
This work     & \cmark & Double & \cmark
& \cmark
& $\begin{array}{c}
     \text{Fermion} \\
     \text{Boson} 
\end{array}$
\\ \hline
$\begin{array}{c}
     \text{Nevanlinna and} \\
     \text{Carath\'{e}odory} 
\end{array}$
 & \xmark & Extended  & \cmark 
  &${\begin{array}{c}
     \text{\cmark}\text{ if clean}\\
     \text{\xmark}\text{ if noisy} 
\end{array}}$  & Fermion
\\ \hline
MaxEnt      & \cmark & Double & \xmark
& \cmark & $\begin{array}{c}
     \text{Fermion} \\
     \text{Boson} 
\end{array}$
\\ \hline
Pad\'{e}     & \xmark & Extended
& \cmark
& \xmark & $\begin{array}{c}
     \text{Fermion} \\
     \text{Boson} 
\end{array}$
\\ 
\hline
\hline
\end{tabular}
\label{table:comparison}
\caption{Comparison of approaches for analytic continuation. Double means double precision arithmetic operations, and extended means extended precision arithmetic operations (requiring 128 bits or larger).}
\end{table*}

This article is organized as follows. In \cref{sec:theory}, we introduce the theory of the PES method. After giving a brief discussion on the setup of the analytic continuation problem in \cref{subsec:setup}, we explain in detail each of the three steps of the PES method: the pre-processing step using semidefinite projection in \cref{subsec:proj}, the pole-estimation step using AAA algorithm in \cref{subsec:AAA} and the SDR fitting step using SDR and bi-level optimization in  \cref{subsec:SDRFit}. We summarize the PES method in \cref{subsec:summary}. The numerical results are presented in \cref{sec:experiments}.
With the Hubbard dimer example, we compare the results of our methods with the Nevanlinna, Carath\'{e}odory and MaxEnt methods, and show that the PES method is both noise-robust and efficient for sharp features, and recovers both diagonal and off-diagonal entries. We also demonstrate the necessity of the pole estimation step using this model. In \cref{subsec:bosonic}, we show that the PES method is also  applicable to bosonic response functions and has a much better performance compared to MaxEnt. In \cref{subsec:bandstructure}, in order to demonstrate the strength of the PES method, we conduct analytic continuations for the band structure of solids with hundreds of orbitals. 

\section{Theory}
\label{sec:theory}
\subsection{Analytic structure of Green's functions}
\label{subsec:setup}

Let $\hat c_i,\hat c_i^{\dagger}$ be the  annihilation and creation operator for the $i$-th orbital, $i=1,\cdots,N_{\text{orb}}$, where $N_{\text{orb}}$ is the number of orbitals.
In the Lehmann representation, a Green's function can be expressed as follows (see e.g.,~\cite[Chapter 5.2]{NegeleOrland1988}): 
\begin{equation}
    \mathbb G_{ij}(z)=\frac{1}{Z}\sum_{r,s}\frac{\left\langle \Psi_s\right|\hat c_i\left|\Psi_r\right\rangle\left\langle \Psi_r\right|\hat c_j^{\dagger}\left|\Psi_s\right\rangle}{z+E_s-E_r}(\mathrm e^{-\beta E_s}\mp\mathrm e^{-\beta E_r}).
    \label{eq:greens}
\end{equation}
Here  $\beta$ is the inverse temperature. $\left|\Psi_s\right\rangle$ is the $s$-th eigenfunction of the Hamiltonian $\hat H$ with energy $E_s$, i.e.,
$$
\hat H\left|\Psi_s\right\rangle = E_s\left|\Psi_s\right\rangle,\quad s = 1,2,\cdots,N_{\mathrm S},
$$
where $N_{\mathrm S} = 2^{N_{\text{orb}}}$, $\hat H$ is the many-body Hamiltonian 
 and $Z=\sum_{s=1}^{N_{\mathrm S}} \mathrm e^{-\beta E_s}$ is the partition function.  The negative sign corresponds to  bosons, while the positive sign corresponds to fermions.

Since 
$\mathbb G_{ij}(z)$  is defined for $i,j\in\{1,\cdots N_{\text{orb}}\}$, the Green's function $\mathbb G$ can be viewed as a $N_{\text{orb}}\times N_{\text{orb}}$ matrix. In other words, we have $\mathbb G(z)=\sum_{r,s}\dfrac{\mathbb X_{r,s}}{z-\lambda_{r,s}}$
where $\lambda_{r,s}=E_r-E_s\in\mathbb{R}$ and $\mathbb X_{r,s}=c_{r,s}v_{r,s}v_{r,s}^{\dagger}$
is a rank-1 $N_{\text{orb}}\times N_{\text{orb}}$ matrix, in which
\begin{equation}
    c_{r,s}=\frac{\mathrm e^{-\beta E_s}\mp \mathrm e^{-\beta E_r}}{Z},\quad v_{r,s}(i)=\left\langle \Psi_r\right|\hat c_i^{\dagger}\left|\Psi_s\right\rangle.
\end{equation}
Here $v_{r,s}(i)$ is the $i$-th component of $v_{r,s}$.
Note that for fermionic systems, $c_{r,s}$ is always nonnegative. For bosonic systems, we have $c_{r,s}> 0$ if $E_r>E_s$, or equivalently  $\lambda_{r,s}>0$; while $c_{r,s}< 0$ if $E_r<E_s$, or equivalently  $\lambda_{r,s}<0$.

For orthogonal orbitals, the creation and annihilation operators satisfy the canonical relation $[\hat{c}_i,\hat{c}_j^{\dagger}]_{\mp} =\hat{c}_i\hat{c}_j^{\dagger}\mp \hat{c}_j^{\dagger}\hat{c}_i= \delta_{ij}$. 
As a result,
the matrices $\mathbb X_{r,s}$ have the following property:
\begin{equation}
    \begin{aligned}
    &\sum_{r,s} \left(\mathbb X_{r,s}\right)_{ij}\\ =  & 
    \frac{1}{Z}\sum_{r,s} \left\langle \Psi_s\right|\hat c_i\left|\Psi_r\right\rangle\left\langle \Psi_r\right|\hat c_j^{\dagger}\left|\Psi_s\right\rangle \left(\mathrm e^{-\beta E_s}\mp\mathrm e^{-\beta E_r}\right) \\
 =    &  \frac{1}{Z}\sum_{s} \left\langle \Psi_s\left|\hat c_i\left(\sum_r\left|\Psi_r\right\rangle\left\langle \Psi_r\right|\right)\hat c_j^{\dagger}\right|\Psi_s\right\rangle \mathrm e^{-\beta E_s}\\
      \mp & \frac{1}{Z}
    \sum_{r} \left\langle \Psi_r\left|\hat c_j^{\dagger}\left(\sum_s\left|\Psi_s\right\rangle\left\langle \Psi_s\right|\right)\hat c_i\right|\Psi_r\right\rangle\mathrm 
    e^{-\beta E_r} \\
      =&\frac{1}{Z}\sum_{s} \left\langle \Psi_s\left|\hat c_i\hat c_j^{\dagger}\mp  \hat c_j^{\dagger}\hat c_i\right|\Psi_s\right\rangle \mathrm e^{-\beta E_s} = \delta_{ij}.
    \end{aligned}
    \end{equation}
This indicates  the sum rule $ \sum_{r,s} \mathbb X_{r,s} = \mathbb I_{N_{\text{orb}}}$,
where $\mathbb I_{N_{\text{orb}}}$ is the $N_{\text{orb}}\times N_{\text{orb}}$ identity matrix. For the rest of this article, let us compress the indices $r,s$ into $l$, and let $N_p$ be the number of matrices $\mathbb X_{r,s}$ that are nonzero, i.e., the number of poles $\lambda_{r,s}$ that actually contribute to the Green's function. In this way, the Green's function $\mathbb G(z)$ can be written as
\begin{equation}
    \mathbb G(z)=\sum_{l=1}^{N_p}\frac{\mathbb{X}_l}{z-\lambda_l},
    \label{eq:Greensfinal}
\end{equation}
where $\mathbb X_l$ satisfies  the rank-1 semidefinite condition:
\begin{equation}
    \begin{array}{cc}\text{For bosons:} & 
    \begin{array}{cc} 
        \operatorname{sign}(\lambda_l)\cdot\mathbb{X}_l \text{ is an } N_{\text{orb}}\times N_{\text{orb}} \\
         \text{rank-1 positive semidefinite  matrix;} 
    \end{array} 
    \end{array} 
    \label{eq:bosonSDconditions}
\end{equation}
\begin{equation}
    \begin{array}{cc}
   \text{For fermions:} &   \begin{array}{cc}
     \mathbb{X}_l \text{ is an } N_{\text{orb}}\times N_{\text{orb}} \\
         \text{rank-1  positive semidefinite  matrix;} 
    \end{array}
    \end{array}
    \label{eq:fermionSDconditions}
\end{equation}
and the sum rule
\begin{equation}
    \sum_{l=1}^{N_p} \mathbb X_{l} = \mathbb I_{N_{\text{orb}}}.
    \label{eq:normcondition}
\end{equation}

The function $\mathbb G(z)$ is defined on $\{0\}\bigcup \mathbb C\backslash\mathbb R$, which excludes the real axis. The Matsubara Green's function $\mathbb G^{\mathrm M}(\omega)$ and the retarded  Green's function $\mathbb G^{\mathrm R}(\omega)$ 
share the same formula  $\mathbb G(z)$ (\cref{eq:greens} and \cref{eq:Greensfinal}) in the following way:
\begin{itemize}
    \item When $z=\mathrm{i}\omega_n\in\mathrm{i}\mathbb{R}$ is on the imaginary axis,  $\mathbb{G}^{\mathrm M}(\omega_n)=\mathbb{G}(\mathrm{i}\omega_n)$ is  the Matsubara (or imaginary frequency) 
    Green's function, $\omega_n$ is called the Matsubara frequency. For fermionic systems, $\mathrm i\omega_n\in\mathrm{i}\frac{2\mathbb{Z}+1}{\beta}$; for bosonic systems,  $\mathrm i\omega_n\in\mathrm{i}\frac{2\mathbb{Z}}{\beta}$.
    \item Since $\mathbb G(z)$ has poles on the real axis, the real-time (or real-frequency) Green's function could only be evaluated at positions infinitesimally close to the real axis. Let $\eta$ be a positive infinitesimal number. When $z = \omega+\mathrm i\eta$, $\mathbb{G}^{\mathrm R}(\omega)=\mathbb{G}(\omega+\mathrm i\eta)$ is the retarded Green's function.
\end{itemize}
The spectral function $A(\omega)$ is defined from $\mathbb{G}^{\mathrm R}(\omega)$
as follows:
\begin{equation}
    A(\omega) = -\frac{1}{\pi} \operatorname{Im}\left({\operatorname{Tr}\left(\mathbb G^{\mathrm{R }} (\omega)\right)}\right) 
\end{equation}
The spectral function contains information of the excitation spectra in quantum systems.
In practice, the retarded Green's function and the spectral function are evaluated with $\eta$ a very small positive number.

 Green's functions of fermionic systems are closely related to the matrix-valued Herglotz-Nevanlinna functions (see e.g., \cite{BelyiHassiSnooEtAl2006}). A matrix-valued function $\mathbb N(z)$ is said to be Herglotz-Nevanlinna if $\mathbb N: \mathbb{C}_{+} \rightarrow \mathbb{C}^{N_{\text{orb}}\times N_{\text{orb}}}$ is analytic and $\operatorname{Im}(\mathbb N(z))$ is a positive semidefinite matrix for $z \in \mathbb{C}_{+}$. Here $\mathbb{C}_{+}$ is the open complex upper half-plane, $\mathbb{C}^{N_{\text{orb}}\times N_{\text{orb}}}$ is the set of $N_{\text{orb}}\times N_{\text{orb}}$ matrices with entries in $\mathbb{C}$, and  $\operatorname{Im}(\mathbb N(z))$  is the imaginary part of $\mathbb N(z)$. The matrix-valued Herglotz-Nevanlinna functions admit the following integral representation, (see e.g., \cite[Theorem 5.4]{GesztesyTsekanovskii2000}  for proof):

\begin{equation}
\mathbb N(z)=Mz+B+\int_{\mathbb{R}}\left(\frac{1}{t-z}-\frac{t}{1+t^{2}}\right) \mathrm d \Sigma(t), \text{ } z \in \mathbb{C}_+,
\label{eq:nevanlinna}
\end{equation}
where $M,B\in \mathbb{C}^{N_{\text{orb}}\times N_{\text{orb}}}$, $M$ is positive semidefinite, $B$ is Hermitian, and $\Sigma(t)$ is a nondecreasing matrix-valued function on $\mathbb{R}$ such that $\int_{\mathbb R}\frac{u^{\dagger}(\mathrm d \Sigma(t))u}{1+t^2}<\infty$ for any vector $u\in\mathbb C^{N_{\text{orb}}}$. This condition on $\Sigma(t)$ ensures that 
the above integration in \cref{eq:nevanlinna} is well-defined. For the detailed
mathematical theory of Herglotz-Nevanlinna functions, we refer the readers to
 \cite[Chapter 2]{ShohatTamarkin1950} and \cite[Chapter 3]{Akhiezer2020}. 

From \cref{eq:nevanlinna}, we may set
\begin{equation}
   \mathrm d\Sigma(t) = \sum_{l=1}^{N_p} \mathbb X_l\delta (t-\lambda_l),\quad M = 0,  \quad B =\int_{\mathbb R} \frac{t}{1+t^2}\mathrm d\Sigma(t),
\end{equation}
then
\begin{equation}
   \mathbb  N(z) =  \int_{\mathbb{R}}\frac{\mathrm d \Sigma(t)}{t-z} = \sum_{l=1}^{N_p}\frac{\mathbb X_l}{\lambda_l-z}=-\mathbb G(z), \quad z \in \mathbb{C}_+.
\end{equation}
In other words, $-\mathbb G(z)$ is a matrix-valued Herglotz-Nevanlinna function. This is the mathematical foundation of Nevanlinna \cite{FeiYehGull2021} and Carath\'{e}odory continuation \cite{FeiYehZgidEtAl2021} methods.

The causal space for Green's functions of fermionic systems is a subset of the Herglotz-Nevanlinna function class denoted by $\mathcal{S}_{\mathrm{F}}$:
\begin{equation}
    \mathcal{S}_{\mathrm{F}} = \left\{
         \mathbb G
    \left|
     \begin{array}{c}
        \mathbb G(z)=\sum_{l=1}^{N_p}\frac{\mathbb{X}_l}{z-\lambda_l},  \text{ for some } N_p\in\mathbb Z,
       \\\text{ for some } \lambda_l  \in\mathbb{R}, \text{ and for } \mathbb{X}_l  \text{ satisfying}\\\text{ fermionic rank-1 semidefinite}
     \text{  conditions}\\\text{ \cref{eq:fermionSDconditions} and}\text{ the sum rule \cref{eq:normcondition}}. 
    \end{array}
  \right.\right\}.
  \label{eq:S-def}
\end{equation}
Similarly, the causal space of Green's functions of bosonic systems is
\begin{equation}
    \mathcal{S}_{\mathrm{B}} = \left\{
         \mathbb G
    \left|
     \begin{array}{c}
        \mathbb G(z)=\sum_{l=1}^{N_p}\frac{\mathbb{X}_l}{z-\lambda_l},  \text{ for some } N_p\in\mathbb Z,
       \\\text{ for some } \lambda_l  \in\mathbb{R}, \text{ and for } \mathbb{X}_l  \text{ satisfying}\\\text{ fermionic rank-1 semidefinite}
     \text{  conditions}\\\text{ \cref{eq:bosonSDconditions} and}\text{ the sum rule \cref{eq:normcondition}}. 
    \end{array}
  \right.\right\}.
  \label{eq:S+def}
\end{equation}
In the definition of $S_{\pm}$, the rank-1 condition in $\mathbb X_l$ can be dropped,  because any semi-definite matrix can be represented as the sum of several rank-1 semi-definite matrices. Degenerate excitations can be treated similarly by allowing the values of some $\lambda_l$'s to be the same. This is the mathematical foundation of the semidefinite relaxation fitting~\cite{Mejuto-ZaeraZepeda-NunezLindseyEtAl2020}.

Now we are ready to introduce the setup for analytic continuation problems of Matsubara data. Given several (possibly noisy) Matsubara data $\mathbb G_n \approx\mathbb{G}(\mathrm i\omega_n)$ for $n=1,2,\cdots,N_w$, our goal is to obtain the fitting of these data into the following analytic form $\mathbb G(z)=\sum_{l=1}^{N_p}\frac{\mathbb{X}_l}{z-\lambda_l}$,
where $\lambda_l\in\mathbb{R}$ and $\mathbb{X}_l$ is a $N_{\text{orb}}\times N_{\text{orb}}$ rank-1 semidefinite matrix, and the matrices $\left\{\mathbb{X}_l\right\}_{l=1}^{N_p}$  satisfies  the sum rule $\sum_l \mathbb{X}_l = \mathbb I_{N_{\text{orb}}}$. For fermionic systems, $\mathbb{X}_l$ is positive semidefinite, while for bosonic cases, $\operatorname{sign}(\lambda_l)\cdot\mathbb{X}_l$ is positive semidefinite.

In summary, we would like to obtain the poles $\lambda_l$, the semidefinite  matrices $\mathbb{X}_l$, and the number of poles $N_p$ which we may not have \textit{a priori} knowledge. 
From such information we can infer the spectral function, as well as other diagonal and off-diagonal entries of the Green's function.

Finally, we would like to remark that the Green's functions are only one type of response functions. Other response functions commonly considered, such as the charge, magnetic, or superconducting susceptibilities (see e.g., \cite[Chapter 5.5]{NegeleOrland1988}) admit the same formula \cref{eq:Greensfinal}, in which $\mathbb X_l$ also satisfies the bosonic/fermionic semidefinite conditions (see \cref{eq:bosonSDconditions,eq:fermionSDconditions}). The only difference is that  $\mathbb X_l$ may satisfy a different set of sum rules. In other words, the  analytic structure of Green's function and other response functions are similar, therefore the same methodology of analytic continuation applies.

\subsection{Step 1: Projection of the noisy data to the causal space}
\label{subsec:proj}

The Matsubara data often contains unphysical noise, i.e., $\mathbb{G}^{\mathrm M}(\mathrm i\omega_n)$ cannot be expressed as a matrix-value function in the set $\mathcal S_{\mathrm{F}/\mathrm{B}}$. Therefore the first step of our algorithm is to project the noisy data into the causal space. 
This can be achieved efficiently using semidefinite programming.
For simplicity, let us choose a fine uniform grid on the real-axis
\begin{equation}
 x_m = -\Lambda_x + mh_x,\quad h_x = \frac{2C}{M},\quad m = 0,\cdots,M,
 \label{eqn:realmesh}
\end{equation}
where $\Lambda_x$ is the real-axis cutoff, and $h_x$ is the grid size.
Our goal is to fit the Matsubara data $\mathbb G_n$ via the following form:
\begin{equation}
    \mathbb G_n \approx \sum_{m=0}^M \frac{\mathbb P_m}{\mathrm i\omega_n - x_m},\quad n =1,\cdots N_w,
\end{equation}
where $\{\mathbb P_m\}$ are semidefinite matrices. 
The objective function $\mathcal E_{\mathrm{proj}}\left(\{\mathbb P_m\}_{m=0}^M\right)$ is defined as 
\begin{equation}
 \mathcal E_{\mathrm{proj}}\left(\{\mathbb P_m\}_{m=0}^M\right)=\left(\sum_{n=1}^{N_w}\left\|\mathbb G_n - \sum_{m=0}^M\frac{\mathbb P_m}{\mathrm{i}\omega_n-x_m}\right\|^2_{F}\right)^{1/2}. \label{eq:projerr}
\end{equation}
Here $\norm{\cdot}_F$ is the Frobenius norm. The solutions of the projection $\mathbb P_m^{\mathrm{proj}}$ is obtained by solving the following optimization problem:
\begin{equation}
\begin{array}{c}
  \{\mathbb P_m^{\mathrm{proj}}\}  =   \arg\min_{\{\mathbb P_m\}}\mathcal E_{\mathrm{proj}}\left(\{\mathbb P_m\}_{m=1}^M\right)\\
\text{ subject to }\left\{\begin{array}{l}
\text{i) semidefinite constraint: }\\ \quad \left\{\begin{array}{rr}
   \mathbb{P}_m\succeq 0  &  \text{(Fermions)},\\
    \operatorname{sign}(x_m)\cdot \mathbb{P}_m\succeq 0   & \text{(Bosons case)}; 
\end{array}   \right. \\
     \text{ ii) sum rules: } \\
     \quad \sum_{m}\mathbb P_m = \mathbb I_{N_{\text{orb}}}.
\end{array}\right.
\end{array}
\label{eq:SDRproj1}
\end{equation}
This is a convex optimization problem, and can be solved efficiently using software packages such as CVX \cite{GrantBoyd2014}.
If we are performing analytic continuation   of other correlation functions, $\{\mathbb P_m\}_{m=1}^M$ may be subject to a different sum rule.

For scalar-valued fermionic Green's function, the positive semidefinite condition $\mathbb P_m\succeq 0$ becomes the non-negativity condition $\mathbb P_m\geq 0$. 
Therefore the sum rule $\sum_m \mathbb P_m = 1$ can also be written as $\sum_m |\mathbb P_m| = 1$, i.e., an $\ell^1$-norm constraint on the vector $(\mathbb P_1,\cdots,\mathbb P_m)$. The least squares problem with a $\ell^1$-norm constraint is similar to the well-known Least Absolute Selection and Shrinkage Operator (LASSO) problem~\cite{Tibshirani1996}, which favors solutions with a sparse structure (see also \cite{CandesRombergTao2006}). 
This agrees with our numerical observation that the solution $\mathbb P_m$ often has relatively few nonzero entries.
For matrix-valued Green's function, there is a similar mechanism that induces the sparsity of solution. The natural generalization of the $\ell^1$ norm is the nuclear norm $\|\cdot\|_*$, defined as $ \|A\|_{*} = \sqrt{\operatorname{Tr}(A^{\dagger}A)} = \sum_k\sigma_k(A)
$, where $\sigma_k(A)$ is the $k$-th largest singular value for $A$. Note that each $\mathbb P_m$ is a positive semidefinite matrix, therefore the $k$-th largest singular value $\sigma_k(\mathbb P_m)$  is equal to the $k$-th largest eigenvalue $\lambda_k(\mathbb P_m)$. Therefore, we have $\sum_m\|\mathbb P_m\|_* = \sum_m\sum_k\sigma_k(\mathbb P_m)= \sum_m\sum_k\lambda_k(\mathbb P_m)=\sum_m\operatorname{Tr}(\mathbb P_m) = \operatorname{Tr}(\sum_m\mathbb P_m) $. Combined with the sum rule $\sum_m \mathbb P_m = \mathbb I_{N_{\operatorname{orb}}}$, we can see that the matrices $\{\mathbb P_m\}$'s are enforced to satisfy the nuclear norm constraint $\sum_m\|\mathbb P_m\|_* = N_{\operatorname{orb}}$. The nuclear norm is also a sparsity-inducing property (see e.g., \cite{BachJenattonMairal2012}).

After obtaining $\mathbb P_m^{\operatorname{proj}}$, we can construct the projected Matsubara data $\mathbb G_n^{\mathrm{proj}}$:
\begin{equation}
    \mathbb G_n^{\mathrm{proj}} =  \sum_{m=0}^M \frac{\mathbb P_m^{\mathrm{proj}}}{\mathrm i\omega_n - x_m},\quad n =1,\cdots N_w,
    \label{eq:reconstruct}
\end{equation}
This projection step can be used to improve  the robustness of other analytic continuation methods, and therefore {is of} independent interest. 
For instance, the Nevanlinna analytic continuation method requires the Pick matrix to be positive semidefinite (see \cite{FeiYehGull2021} for details). Numerical results indicate that this criterion  can be violated when very small perturbations are applied to the Matsubara data. 
Given that noise is inevitable in many Green's function solvers (notably, quantum Monte Carlo solvers), the application range of the Nevanlinna analytic continuation is thus significantly limited by the nature of the noise. 
Our numerical results suggest that when the Nevanlinna method is applied to the projected noisy Matsubara data (for diagonal entries), the quality of the analytic continuation is significantly improved. (See \cref{subsec:hubbarddimer} and \cref{subsec:bandstructure} for details.)
Similar improvements are observed on both diagonal and off-diagonal entries of the Green's functions while applying Carath\'{e}odory continuations on the projected data (also see \cref{subsec:hubbarddimer} for details).

\subsection{Step 2:  Estimation of pole locations using the AAA algorithm}
\label{subsec:AAA}

While the projection step can be viewed as an analytic continuation algorithm by itself, the quality of the continuation is constrained by the resolution of the grid on the real axis. 
In practice, the number of grid points (i.e., $M$) in the uniform grid is often too small to accurately resolve the pole locations, but is too large to be directly used in the subsequent semidefinite relaxation (SDR) step to be detailed in \cref{subsec:SDRFit}. 
Furthermore, the loss function in the SDR step is highly non-convex, and the optimization with respect to this loss function requires a proper initial guess. These considerations lead to the second step of the algorithm for estimating the locations of a relatively small number of poles.

We use the adaptive Antoulas--Anderson (AAA) algorithm \cite{NakatsukasaSeteTrefethen2018} for the pole estimation, which  is available as a subroutine in the Chebfun package \cite{DriscollHaleTrefethen2014}. The AAA algorithm is  able to obtain the poles of a scalar complex-valued function $g(z)$. In  the context of analytic continuation, the scalar function $g(z)$ could either be each diagonal entry $\mathbb G_{ii}(z) (i=1,\cdots,N_{\text{orb}})$ of the Green's function, or the trace $\operatorname{Tr}\left(\mathbb{G}(z)\right)$. In practice, we find that performing the pole estimation on each diagonal entry separately often gives better results.

The AAA algorithm  performs a  rational approximation to a scalar function $g(z)$ using barycentric representations~\cite{BerrutTrefethen2004}:
 \begin{equation}
g(z)\approx \frac{n(z)}{d(z)}=\frac{\sum_{j=1}^{n} \frac{w_{j} g_{j}}{z-z_{j}}} {\sum_{j=1}^{n} \frac{w_{j}}{z-z_{j}}}.
\label{eq:barycentric}
\end{equation}
Here $n(z) =\sum_{j=1}^{n} \dfrac{w_{j} g_{j}}{z-z_{j}}  $ and  $d(z)=\sum_{j=1}^{n} \dfrac{w_{j}}{z-z_{j}}$ are the numerator and denominator part of the barycentric representation, respectively.
Note that the above formulation \cref{eq:barycentric} automatically satisfies that $\frac{n(z_j)}{d(z_j)}=g_j$. 

We briefly describe the AAA algorithm below, and refer readers to Ref.~\cite{NakatsukasaSeteTrefethen2018} for more details. Assume that we are given the values of $g(z)$ on a set $\mathcal{Z}\subset \mathbb C$ of $N$ points (in analytic continuation, $\mathcal{Z}=\{\mathrm i\omega_n, n=1,\cdots,N_w\}$),  the AAA algorithm aims at finding a set of support points $\{z_j\}_{i=1}^n$, which allows us to solve for the weights $\{w_j\}_{j=1}^n$ in \cref{eq:barycentric}. Here $g_j=g(z_j)$  ($j=1,\cdots,n$). Starting from $n=1$, the AAA algorithm gradually expands the set $\{z_j\}_{i=1}^n$ following a greedy algorithm.

At the $(n-1)$-th step ($n=1,2,\cdots$), if the residual
$\mathrm{res}(z)=g(z)-\frac{n(z)}{d(z)}$ is sufficiently small for all $z\in \mathcal{Z}$, then the algorithm terminates. Otherwise, we pick the next support point $z_n$, by choosing  $z\in \mathcal{Z} \backslash\left\{z_{1}, \ldots, z_{n-1}\right\}$ where the residual $\mathrm{res}(z)$  takes its maximum absolute value. 
Let us write $\mathcal{Z} \backslash\left\{z_{1}, \cdots, z_{n}\right\}$ as a column vector 
    ${\boldsymbol Z}^{(n)}:=\left(Z_{1}^{(n)}, \cdots, Z_{N-n}^{(n)}\right)^{T}$ and define 
${\boldsymbol g}^{(n)}=\left(g_1^{(n)},\cdots,g_{N-n}^{(n)}\right)^T := 
    \left(g\left(Z_{1}^{(n)}\right), \cdots, g\left(Z_{N-n}^{(n)}\right)\right)^{T}$.
    Since $g(z)$ is supposed to be approximated by $g(z)\approx n(z)/d(z)$, i.e. $g(z)d(z)\approx n(z)$, we choose the normalized weight $w=(w_1,\cdots,w_n)$ by solving the least square problem:
    \begin{equation}
    \begin{aligned}
    \min_{\| w\|_2=1}&\sum_{i=1}^{N-n} \left|g_{i}^{(n)}d\left(Z_{i}^{(n)}\right)-n\left(Z_{i}^{(n)}\right)\right|^2
        \\
        =&\sum_{i=1}^{N-n} \left|\sum_{j=1}^{n} \frac{w_{j}\left(g_i^{(n)}-g_j\right)}{Z_i^{(n)}-z_{j}}\right|^2.
    \end{aligned}
        \label{eq:aaaopt}
    \end{equation}
The minimization problem of \cref{eq:aaaopt} could be viewed as the following linear algebra problem:
    \begin{equation}
    \min_{\|w\|_{2}=1}\left\|A^{(n)} w\right\|_{2},
    \label{eq:aaaaopt3}
    \end{equation}
where $A^{(n)}$ is 
\begin{equation}
    \begin{aligned}
    A^{(n)}=\left(\begin{array}{ccc}
\frac{g_{1}^{(n)}-g_{1}}{Z_{1}^{(n)}-z_{1}} & \cdots & \frac{g_{1}^{(n)}-g_{n}}{Z_{1}^{(n)}-z_{n}} \\
\vdots & \ddots & \vdots \\
\frac{g_{N-n}^{(n)}-g_{1}}{Z_{N-n}^{(n)}-z_{1}} & \cdots & \frac{g_{N-n}^{(n)}-g_{n}}{Z_{N-n}^{(n)}-z_{n}}
\end{array}\right).
    \end{aligned}
\end{equation}
This optimization problem \cref{eq:aaaaopt3} could be solved by performing a singular value decomposition (SVD) on the matrix $A^{(n)} =U \Sigma V^{\dagger}$, where $U$ is a $(N-n)\times r$ orthogonal matrix, $V$ is a $n\times r$ orthogonal matrix,  $\Sigma=\operatorname{diag}(\sigma_1,\cdots,\sigma_r)$ is a  diagonal matrix ($\sigma_1\geq \cdots \geq \sigma_r$), and $r\leq \min (N-n,n)$ is the rank of $A^{(n)}$. The solution $w$ of \cref{eq:aaaaopt3}
should be taken as the last column of $V$.

After the iteration terminates at the $K$-th step, we have $\{z_1,\cdots,z_K\}$ and its associated weights $\{w_1,\cdots,w_K\}$. We can calculate the zeros of $d(z)$, which serve as the estimated poles of $g(z)$ that we want to obtain, using the following generalized eigenvalue problem \cite{NakatsukasaSeteTrefethen2018}: 
\begin{equation}
\left(\begin{array}{ccccc}
0 & w_{1} & w_{2} & \cdots & w_{K} \\
1 & z_{1} & & & \\
1 & & z_{2} & & \\
\vdots & & & \ddots & \\
1 & & & & z_{K}
\end{array}\right)=\xi\left(\begin{array}{ccccc}
0 & & & & \\
& 1 & & & \\
& & 1 & & \\
& & & \ddots & \\
& & & & 1
\end{array}\right).
\end{equation}
At least two of the eigenvalues in this problem are infinite, and the remaining $(K-1)$
eigenvalues $\{\xi_{a}\}_{a=1}^{K-1}$ are the zeros of $d$, i.e. the complex-valued poles of $g$. 

Since all poles of the Green's function in \cref{eq:Greensfinal} are real-valued, after obtaining $\{\xi_{a}\}_{a=1}^{K-1}$,  we discard the poles far away from the real axis. In other words, we only keep poles with $\abs{\Im \xi_a}\le \varepsilon_p$ for some $\varepsilon_p>0$, and define $\{\lambda_l^{\operatorname{in}}\}_{l=1}^{N_p}$ to be real parts of the remaining poles. 
These $N_p$ poles will be used as the initial guess of the poles in the 
semidefinite relaxation step below.

\subsection{Step 3: Semidefinite relaxation}
\label{subsec:SDRFit}

In the final step, we use numerical optimization to obtain an effective fitting of the Matsubara data in the form of \cref{eq:Greensfinal}. The fitting error is defined as
\begin{equation}
\begin{aligned}
     &\text{Err}\left(\{\lambda_l\}_{l=1}^{N_p},\{\mathbb X_l\}_{l=1}^{N_p}\right)\\
     =&\left(\sum_{n=1}^{N}\left\|{\mathbb G}_n-\mathbb G\left(\mathrm{i}\omega_n;\{\lambda_l,\mathbb X_l\}_{l=1}^{N_p}\right)\right\|_F^2\right)^{1/2}\\
     =&\left(\sum_{n=1}^{N}\left\|
    \mathbb G_n-\sum_{l=1}^{N_p}\frac{\mathbb X_l}{\mathrm{i}\omega_n-\lambda_l}
     \right\|_F^2\right)^{1/2}.
     \label{eq:Err}
\end{aligned}
\end{equation}
The fitting error expressed in \cref{eq:Err} is a highly nonconvex function, and the minimization of this function can frequently be trapped in local minima, which strongly depend on the choice of the initial guess. 

Recall that in \cref{eq:Greensfinal} $\{\mathbb{X}_l\}$ are required to be rank-1 semidefinite matrix.
The semidefinite relaxation  drops this rank-1 constraint. In other words,  $\{\mathbb X_l\}$ (or $\operatorname{sign}(\lambda_l)\cdot\mathbb X_l$) are only required  to be positive semidefinite matrices for the analytic continuation of fermionic (bosonic) Green's function. 

For simplicity, the following discussion focuses on fermionic Green's functions, while the same numerical treatment is applicable to bosonic Green's functions.
The minimization of $\text{Err}\left(\{\lambda_l\}_{l=1}^{N_p},\{\mathbb X_l\succeq 0\}_{l=1}^{N_p}\right)$
can be formulated as a bi-level optimization problem: 
\begin{equation}
 \begin{aligned}
      &\min_{\lambda_   l,\mathbb X_l\succeq 0}
     \text{Err}\left(\{\lambda_l\}_{l=1}^{N_p},\{\mathbb X_l\}_{l=1}^{N_p}\right)\\=& \min_{\lambda_l\in\mathbb{R}}\min_{\mathbb X_l\succeq 0}
     \text{Err}\left(\{\lambda_l\}_{l=1}^{N_p},\{\mathbb X_l\}_{l=1}^{N_p}\right)\\=&\min_{\lambda_l\in\mathbb{R}} \mathcal{E}\left(\{\lambda_l\}_{l=1}^{N_p}\right)
 \end{aligned},
   \end{equation}
   where
   \begin{equation}
       \mathcal{E}\left(\{\lambda_l\}_{l=1}^{N_p}\right)=\min_{\mathbb X_l\succeq 0}
     \text{Err}\left(\{\lambda_l\}_{l=1}^{N_p},\{\mathbb X_l\succeq 0\}_{l=1}^{N_p}\right).
     \label{eq:poleerr}
   \end{equation}
   When the poles $\{\lambda_l\}_{l=1}^{N_p}$ are fixed, the optimization with respect to semidefinite matrices $\{\mathbb X_l\succeq 0\}$ is a convex optimization problem, and could be efficiently solved using software packages such as CVX \cite{GrantBoyd2014}. The optimization of $ \text{Err}\left(\{\lambda_l\}_{l=1}^{N_p},\{\mathbb X_l\succeq 0\}_{l=1}^{N_p}\right)$ is transformed into the optimization of $\mathcal{E}\left(\{\lambda_l\}_{l=1}^{N_p}\right)$, but note that such an optimization over the pole positions is still a nonconvex problem.

  Therefore, we have the following SDR fitting procedure through the bi-level optimization framework: starting from an initial value for poles $\{\lambda_l^{\operatorname{in}}\}_{l=1}^{N_p}$ obtained by the previous pole estimation step, we use numerical optimization technique to minimize $ \mathcal{E}\left(\{\lambda_l\}_{l=1}^{N_p}\right)$, where
   \begin{itemize}
       \item The value of $ \mathcal{E}\left(\{\lambda_l\}_{l=1}^{N_p}\right)$ and the corresponding optimal $\{\mathbb X_l^{\text{opt}}\succeq 0\}$ for fixed poles  $ \{\lambda_l\}_{l=1}^{N_p}$ are found by a convex optimization solver. 
       \item 
       Since $ \mathcal{E}\left(\{\lambda_l\}_{l=1}^{N_p}\right) = \text{Err}\left(\{\lambda_l\}_{l=1}^{N_p},\{\mathbb X_l^{\text{opt}}\}_{l=1}^{N_p}\right)$ and the optimal matrices $\{\mathbb X_l^{\text{opt}}\}$ should be regarded as functions of the poles $\{\lambda_l\}$, 
       the partial derivative of $ \mathcal{E}\left(\{\lambda_l\}_{l=1}^{N_p}\right)$ could be written as
\begin{equation}
       \begin{aligned}
\partial_{m}\mathcal{E}
          & =\frac{\partial}{\partial \lambda_m} \text{Err}
          \\&+\sum_{l=1}^{N_p}\sum_{i,j=1}^{N_{\text{orb}}}
        \left.  \frac{\partial \text{ Err}}{\partial \left(\mathbb X_l\right)_{ij}}\right|_{\mathbb X_l^{\text{opt}}}\cdot
          \frac{\partial \left(\mathbb X_l^{\text{opt}}\right)_{ij}}{\partial \lambda_m}
       \end{aligned},
       \end{equation}
The first order optimality condition for solving \cref{eq:poleerr} implies $\left.  \frac{\partial \text{ Err}}{\partial \left(\mathbb X_l\right)_{ij}}\right|_{\mathbb X_l^{\text{opt}}}=0$. Such an evaluation of the derivative quantities is similar to the treatment in the Hellmann-Feynman's theorem  \cite{Feynman1939}.
  Therefore we have
         \begin{equation}
           \begin{aligned}
         \partial_{m}&\mathcal{E}\left(\{\lambda_l\}_{l=1}^{N_p}\right)
 =\frac{\partial}{\partial \lambda_m} \text{Err}\left(\{\lambda_l\}_{l=1}^{N_p},\{\mathbb X_l\}_{l=1}^{N_p}\right)\Big\vert_{\mathbb X_l=\mathbb X_l^{\text{opt}}}\\=&-\frac{\sum\limits_{n=1}^N\sum\limits_{i,j=1}^{N_{\text{orb}}}\operatorname{Re}\left(\left(
     (\mathbb G_n)_{ij}-\sum\limits_{l} \frac{(\mathbb X_l^{\text{opt}})_{ij}}{\mathrm{i}\omega_n-\lambda_l}
     \right)\frac{(\mathbb X_m^{\text{opt}})_{ij}^*}{(\mathrm{i}\omega_n+\lambda_m)^2}
     \right)}{\sqrt{\mathcal{E}\left(\{\lambda_l\}_{l=1}^{N_p}\right)}}.
           \end{aligned}
           \label{eq:polegradient}
         \end{equation}
   \end{itemize}
   Therefore, both the function value and gradient of $\mathcal{E}$ can be efficiently evaluated. The optimization of  $ \mathcal{E}\left(\{\lambda_l\}_{l=1}^{N_p}\right)$ could be conducted with standard gradient-based optimization solvers, such as the Broyden-Fletcher-Goldfarb-Shanno (BFGS)  method (see e.g. \cite{NocedalWright1999}).
   
Note that the SDR fitting step enforces that the Green's functions must be in the causal space, and is thus robust to unphysical noises. 
Since the projection step may introduce biases especially when the uniform mesh in \cref{eqn:realmesh} is relatively coarse, we find that the performance of SDR fitting is often better when applied to the original noisy data, instead of the data obtained after the projection. 

\subsection{Summary of the PES method}
\label{subsec:summary}
We  summarize the PES method  in \cref{alg:PES} below.

\begin{algorithm}[H]
\caption{Projection-Estimation-Semidefinite relaxation (PES) method}
\label{alg:PES}
\begin{algorithmic}[1]
\STATE{\textbf{Parameters:} real axis cutoff $\Lambda_x$  and mesh size $h_x$ in the projection step; imaginary axis cutoff $\varepsilon_p$ in the pole estimation step}.
\STATE{\textbf{Input:} Matsubara data $\left\{\mathbb G_n\right\}_{n=1}^{N_w}$, where $\mathbb G_n$ is an  $N_{\text{orb}}\times N_{\text{orb}}$ matrix.}
\STATE{\textbf{Output:} Poles $\left\{\lambda_l\right\}^{N_p}_{l=1}$, weights $\{\mathbb X_l\}^{N_p}_{l=1}$.}
\STATE {\textbf{Projection}} of the noisy data $\left\{\mathbb G_n\right\}_{n=1}^{N_w}$: form a uniform mesh in \cref{eqn:realmesh},  minimize $\mathcal E_{\mathrm{proj}}\left(\{\mathbb P_m\}_{m=1}^M\right)$ in \cref{eq:SDRproj1}, and then obtain the projected data  $\left\{\mathbb G_n^{\text{proj}}\right\}_{n=1}^{N_w}$ using \cref{eq:reconstruct}.
\STATE {\textbf{Estimation}} of the pole locations using the projected data $\left\{\mathbb G_n^{\text{proj}}\right\}_{n=1}^{N_w}$ via the AAA algorithm. The output is $\left\{\lambda_l^{\text{in}}\right\}_{l=1}^{N_p}$.
\STATE {\textbf{Semidefinite relaxation fitting}} through bi-level optimization: with the initial poles $\left\{\lambda_l^{\text{in}}\right\}_{l=1}^{N_p}$, we can conduct SDR fitting by minimizing $ \mathcal{E}\left(\{\lambda_l\}_{l=1}^{N_p}\right)$ in \cref{eq:poleerr} to obtain the final poles $\left\{\lambda_l\right\}_{l=1}^{N_p}$ and weights $\left\{\mathbb X_l\right\}_{l=1}^{N_p}$.
\end{algorithmic}
\end{algorithm}

\noindent\textit{Computational cost:} Both the projection and the semidefinite relaxation steps require the solution of convex optimization problems that can be reformulated into semidefinite programming (SDP) problems. 
The cost for solving the convex optimization problems can depend on the SDP reformulation, and the choice of algorithms. 
The default solver in the CVX software package is SDTP3 \cite{TohToddTutuncu2012}, which implements the interior point method with a primal-dual path-following strategy \cite{Mehrotra1992} to solve SDP problems. 
In the primal-dual interior point method, the cost is dominated by solving the Newton equation to obtain the primal-dual search direction. 
The unknown variables are $N_v$ positive semidefinite matrices each of size $N_d\times N_d$, and the total number of variables is $N_v N_d^2$. Therefore the cost of solving Newton's equation can be as large as $\Or(N_v^3 N_d^6)$, and there is an additional cost in forming the matrices in Newton's equation (see e.g.,  \cite{VandenbergheBoyd1996}). While the cost of solving Newton's equation may be reduced using preconditioned and iterative solvers in certain cases~\cite{JiangSunToh2012}, in general, the solution the SDP problem can become expensive when both $N_v$ and $N_d$ are large. 

In the projection step, $N_v=M$ is usually large. 
Note that the pole estimation step only uses the diagonal entries of the Green's function, and it is sufficient to only conduct projection of these diagonal entries.
If $M$ is very large, we may further reduce the cost by performing projection for each diagonal entry of the Green's functions separately.
The semidefinite relaxation fitting requires the solution of $N_v=N_p$ matrices each of size $N_d=N_{\text{orb}}$. Therefore it is important for the pole estimation step to obtain a relatively small number of poles. 
Furthermore, if  the non-diagonal information of the Green's function is not needed (for example if the goal is to calculate the spectral function), the cost can be significantly reduced by conducting the semidefinite relaxation step for the diagonal entries only. Compared to solving the SDP problems, the cost of the AAA algorithm for the pole estimation is usually negligible.

\vspace{1em}
\noindent\textit{Noiseless data:} If the data is noiseless or if the Green's functions are causal by construction (e.g., Green's functions are approximated via causal diagrammatic methods),
we can skip the projection step, and conduct estimation and SDR fitting directly.  Furthermore, since the projection is only performed on a finite grid along the real axis, the pre-processing step may become detrimental to the accuracy when the grid size is relatively small.

\vspace{1em}
\noindent\textit{Parameters in AAA:} Numerical results suggest that the final result is often insensitive to the choice of the parameter $\varepsilon_p$. For noisy data, the Lawson refinement step \cite{NakatsukasaTrefethen2020} (also implemented in Chebfun), which is based on an iteratively reweighted least-squares (IRLS) iteration can also be used to improve the fitting result. Also, the AAA algorithm in Chebfun \cite{DriscollHaleTrefethen2014} could be implemented with or without specifying the number of poles. Prior knowledge on the on the number of peaks in the physical system can be used to control the maximal number of poles obtained by the AAA algorithm, and to enhance the robustness with respect to noise.

\section{Results}
\label{sec:experiments}
In order to demonstrate the strengths of the PES method and give a fair comparison with other methods, we present various results using the Hubbard dimer system and the band structure calculation of different materials. When testing the performances of analytic continuation methods in the presence of noise, we will manually add  multiplicative Gaussian noise to the clean Matsubara data:
$$
\mathbb G_{\text {noisy }}=\mathbb G_{\text {exact }} \cdot\left(1+\sigma N_{\mathbb{C}}(0,1)\right)
$$
Here $N_{\mathbb{C}}(0,1)$ is the complex-valued normal Gaussian noise and $\sigma$ is referred to as the (multiplicative) noise level of the data.
\subsection{Hubbard dimer}
\label{subsec:hubbarddimer}
\subsubsection{Setup of the Hubbard Dimer}
The Hubbard dimer example used in \cite{FeiYehZgidEtAl2021} is a simple but nontrivial test case, since it has multiple sharp features while the true value of the Green's function could be obtained by the exact diagonalization.
It is also a prototypical system for studying the excitation spectra of molecules. There are only four orbitals in the Hubbard dimer, namely $|0\uparrow\rangle,|0\downarrow\rangle,|1\uparrow\rangle, |1\downarrow\rangle$.
 The Hamiltonian is
\begin{equation}
\hat H=\hat H_{0}+\hat H_{1},
\end{equation}
where
\begin{equation}
\begin{aligned}
\hat H_{0}&=-\sum_{\sigma\in\{\uparrow,\downarrow\}} t\left(\hat c_{0 \sigma}^{\dagger} \hat c_{1 \sigma}+
\hat c_{1 \sigma}^{\dagger} \hat c_{0 \sigma}\right)\\
-& \mu\left( \hat n_{0, \uparrow}+\hat n_{0, \downarrow}+\hat n_{1, \uparrow}+\hat n_{1, \downarrow}\right),
\end{aligned}
\end{equation}
\begin{equation}
\begin{aligned}
\hat H_{1}&= (U+U_a)\hat n_{0, \uparrow}\hat n_{0, \downarrow}+(U-U_a)\hat n_{1, \uparrow}\hat n_{1, \downarrow} \\
&-\frac{{U}}{2} \left( \hat n_{0, \uparrow}+\hat n_{0, \downarrow}+\hat n_{1, \uparrow}+\hat n_{1, \downarrow}\right)\\&+ h\left(\hat n_{0 \uparrow}-\hat n_{0 \downarrow}+\hat n_{1 \uparrow}-\hat n_{1 \downarrow}\right)
\\
&+\mu_{a}\left(\hat n_{0 \uparrow}+\hat n_{0 \downarrow}-\hat n_{1 \uparrow}-\hat n_{1 \downarrow}\right)\\&+h_{a}\left(\hat n_{0 \uparrow}-\hat n_{0 \downarrow}-\hat n_{1 \uparrow}+\hat n_{1 \downarrow}\right),
\end{aligned}
\end{equation}
and
\begin{equation}
    \hat n_{i\sigma}=\hat c_{i\sigma}^{\dagger}\hat c_{i\sigma},\quad i=0,1,\quad \sigma=\uparrow,\downarrow.
\end{equation}
The parameters are chosen the same as in \cite{FeiYehZgidEtAl2021}, namely
$ t=1, \mu=0.7, h=0.3, U_{a}=0.5, \mu_{a}=0.2, h_{a}=0.03, \beta = 31.1$ and $U=5$.

\subsubsection{Fitting results, with comparison to other methods}
\begin{figure*}[t]
    \centering
    \includegraphics[width=155mm]{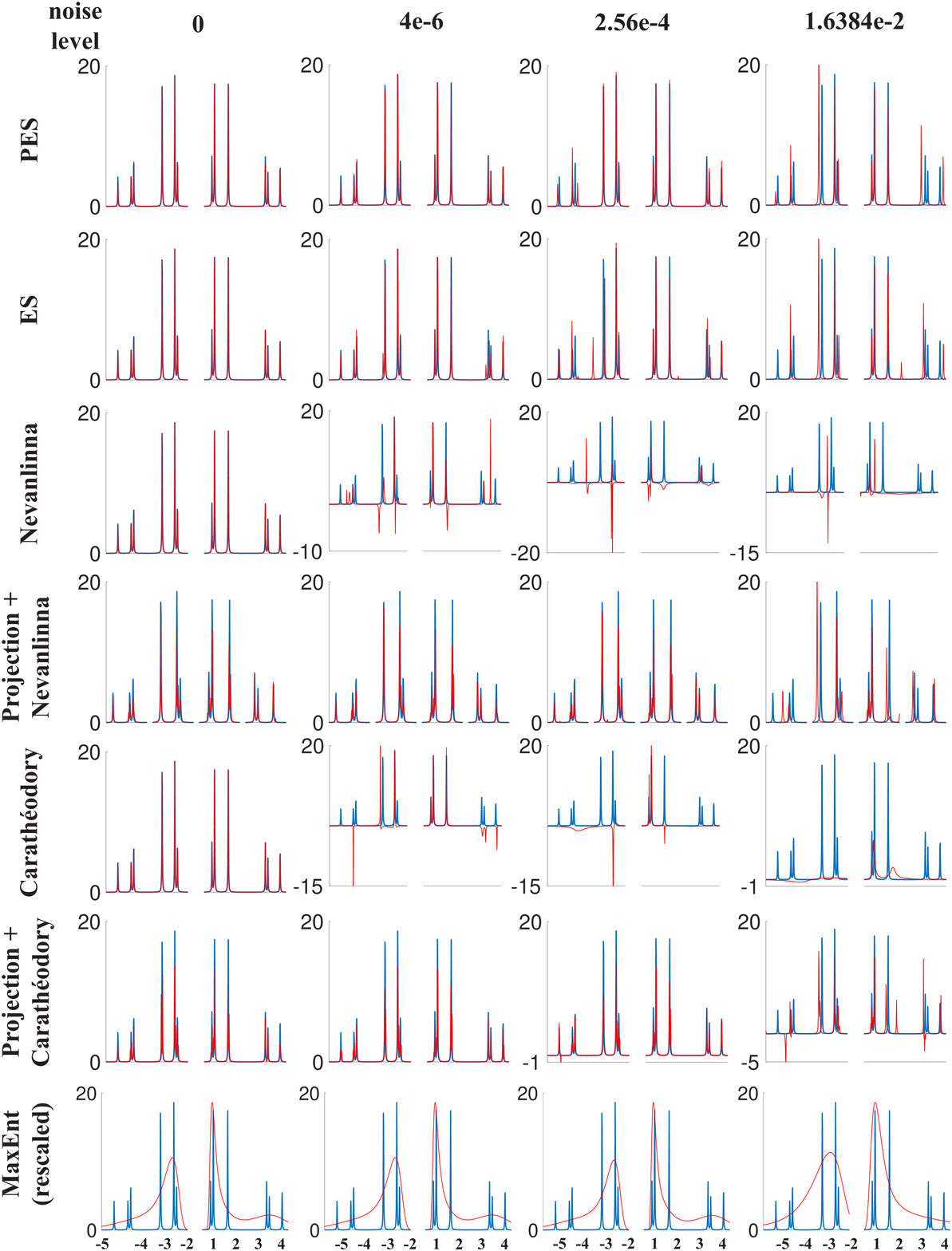}
    \caption{Spectral functions of the Hubbard dimer system obtained from PES, ES, Nevanlinna, Projection+Nevanlinna, Carath\'{e}odory, Projection+Carath\'{e}odory and MaxEnt methods. Blue line: true value. Red line: calculation results from analytic continuation. 
}
    \label{fig:hubbarddimer}
\end{figure*}

Let us first demonstrate the efficiency of the PES method.
We conduct analytic continuation for four sets of Matsubara data with  different noise levels: $\sigma = 0,\sigma =4\times 10^{-6},\sigma =2.56\times 10^{-4}, \sigma =1.6384\times 10^{-2}$.
In \cref{fig:hubbarddimer}, we plot the spectral function (evaluated at $\mathbb R+\mathrm{i}\eta$, $\eta = 0.01$) obtained from analytic continuation with comparison to its true values. The methods that we are comparing with PES are Nevanlinna \cite{FeiYehGull2021}, Carath\'{e}odory \cite{FeiYehZgidEtAl2021} and MaxEnt \cite{LevyLeBlancGull2017}. We also implement the ES method, in which we skip the projection step and conduct the estimation and the SDR fitting directly.
To make it a fair comparison, we also perform Nevanlinna and Carath\'{e}odory continuations on the projected data. Since  the amplitude of MaxEnt spectral functions  are very small compared to the delta peaks in the true spectral functions, we rescale the MaxEnt results for comparison with true values and results from other methods.

Our observation is summarized as follows:
\begin{itemize}
    \item When the Matsubara data is clean (noise-level $\sigma=0$), all methods except the MaxEnt continuation could retrieve all the peaks perfectly. The MaxEnt method can not deal with sharp features. In all cases, MaxEnt can only result in two (and up to three) significantly broadened peaks. 
    \item Both the PES method and the ES method are robust to noise. It can  retrieve almost all  features of the Green's function when the noise  level is small ($\sigma=4\times 10^{-6}$ and $2.56\times 10^{-4}$), and can  retrieve quite a few features when the noise level is relatively large ( $\sigma=1.6384\times 10^{-2}$). Particularly, the band gap could be accurately resolved even in the large noise scenario.
    
    \item For noisy data, PES behaves  better than ES (see noise level $\sigma=4\times 10^{-6}$ and $2.56\times 10^{-4}$), which implies that the projection of noisy data is indeed helpful. For noiseless data ($\sigma=0$), ES  behaves better than PES.
    \item Even when the data is only slightly noisy (noise-level $\sigma=4\times 10^{-6}$), the Nevanlinna and Carath\'{e}odory continuation breaks down. {This is due to the violation of the Pick criterion \cite{FeiYehGull2021}.} The spectral function is no longer nonnegative and multiple features are missing. 
    \item If combined with the projection of data into the causal space, the results of the Nevanlinna and Carath\'{e}odory continuation are greatly improved. However, the quality of Projection+Nevanlinna and Caratheodory is still not as good as PES. This could be seen in all noisy levels.
    What's more, the calculation results are still not guaranteed to be causal, as shown in the results of Projection+Carath\'{e}odory methods for relatively large noise (i.e. $\sigma = 2.56\times 10^{-4}$ and $\sigma=1.6384\times 10^{-2}$).
\end{itemize}

\begin{figure*}
    \centering
    \includegraphics[width = 155mm]{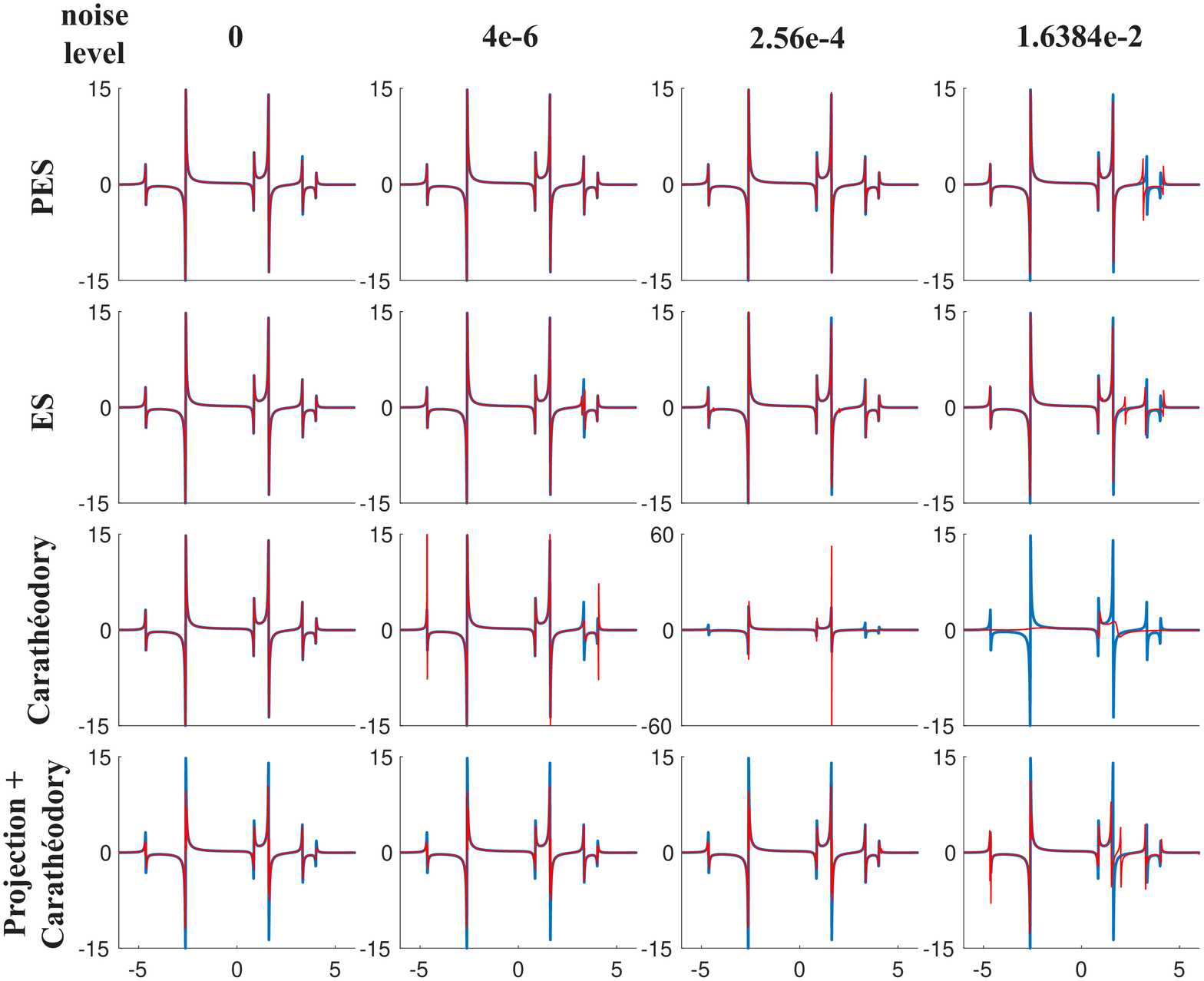}
    \caption{The non-diagonal element $G_{13}$ obtained by different analytic continuation methods. Blue line: true value. Red line: calculation results from analytic continuation.  }
    \label{fig:HubbardG13}
\end{figure*}

Since both  PES method and Carath\'{e}odory continuation are able to obtain non-diagonal entries of the Green's functions, 
let us show the calculation result of the real part of $G_{13}$ (also evaluated at $\mathbb R+\mathrm i\eta$, $\eta = 0.01$)  in \cref{fig:HubbardG13}, in which we compare the performances of PES, ES, Carath\'{e}odory and Projection+Carath\'{e}odory.

 Similar to what we have learned from the results of spectral functions, we have the following observations:
\begin{itemize}
\item The Projection+Carath\'{e}odory does not perform as good as the PES and ES method. This could be seen at all noise levels.   The two high peaks are perfectly retrieved by the PES and ES method but not by Projection+Carath\'{e}odory.
\item At noise-level $\sigma = 1.6384\times 10^{-2}$, the noise is too large and no method could recover all peaks perfectly.
\end{itemize}

With these observations, we conclude that the PES method has the most robust performance, especially for analytic continuation of the matrix-valued noisy Matsubara data.
The causal projection could help cure the noise-sensitivity issue of Nevanlinna and Carath\'eodory continuation, but does not always guarantee a causal result, and not necessarily perform as well as the PES method.

\subsubsection{Importance of the pole estimation}
Now we would like to demonstrate the importance of the pole estimation step. 
In \cite{Mejuto-ZaeraZepeda-NunezLindseyEtAl2020}, the SDR step is implemented in the context of hybridization fitting without the pole estimation. A random initialization of poles could fail to converge towards the global minimum of the loss function (see \cite[Figure 2]{Mejuto-ZaeraZepeda-NunezLindseyEtAl2020}).
In the context of analytic continuation, we also find that a random selection of initial poles will not converge to its true positions. 
Particularly, the high frequency features, i.e. the poles away from zero may not be identified accurately. 

\begin{figure}[htbp]
    \centering
    \includegraphics[width = 85mm]{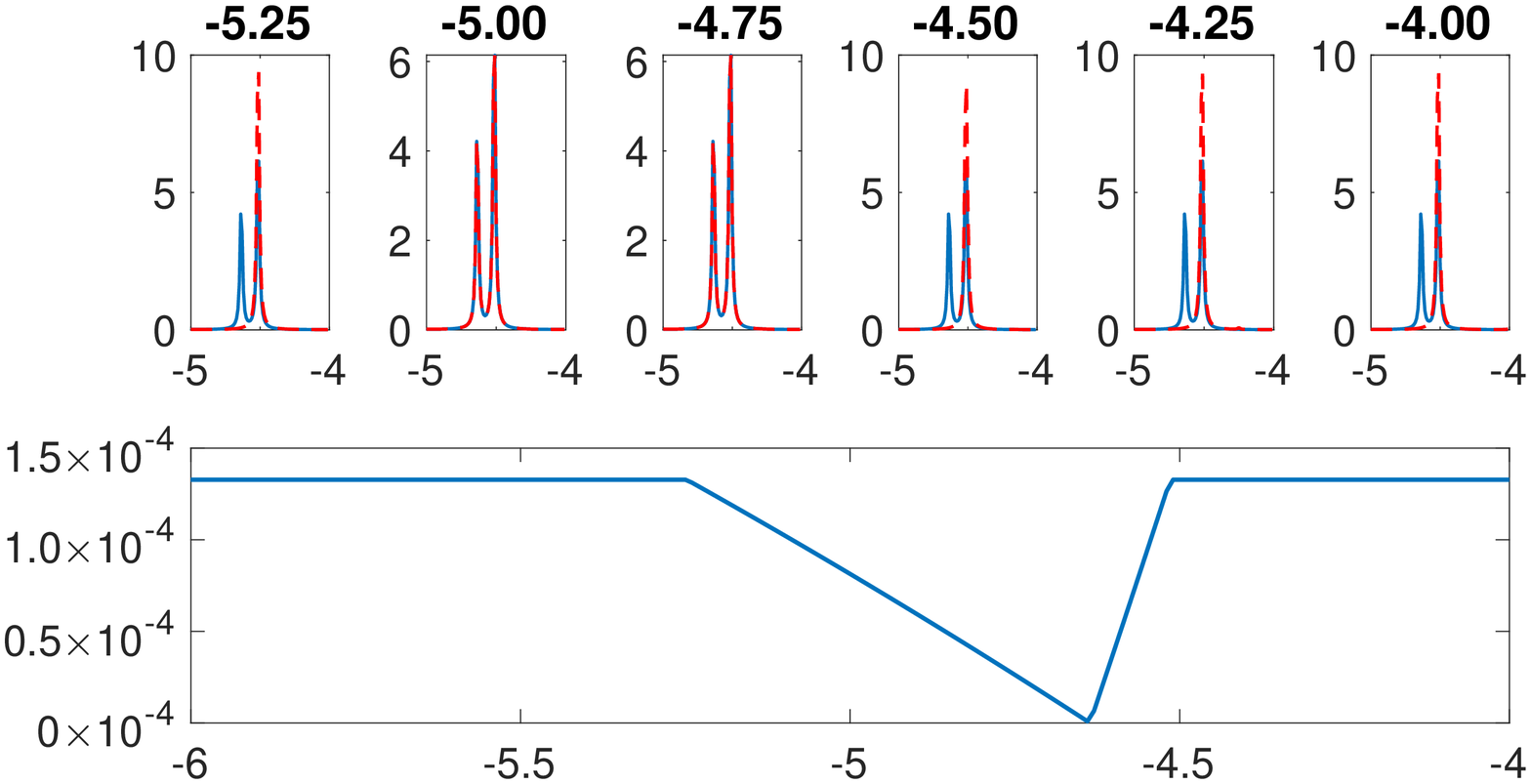}
    \caption{Top panel: calculation results of spectral functions (red dashed lines) with comparison to the true value (blue lines) on the interval $[-5,-4]$. The initial value of $\lambda_1$ are chosen as $\lambda_1^{\text{in}} = -5.25, -5.00, -4.75, -4.50, -4.25$ and $-4.00$. Bottom panel: landscape of  $\mathcal{E}\left(\{\lambda_l\}_{l=1}^{N_p}\right)$ with respect to $\lambda_1$ on the interval $[-6,-4]$. Other poles are kept fixed at their true positions:  $-5.2479$, $-4.5143$, $-3.1957$, $-2.6075$, $-2.4857$, $0.8619$, $0.9837$, $1.6181$, $3.4599$, $3.3381$ and $4.0255$.}
    \label{fig:lambda_1}
\end{figure}

To illustrate the difficulty in the SDR step, we fix the poles to be its true values except for $\lambda_1$. The true value of $\lambda_1$ is  $-4.6361$. We choose its initial value to be $\lambda_1^{\text{in }}= -5.25, -5.00, -4.75, -4.50, -4.25, -4.00$, and then conduct the SDR fitting step. The corresponding result of the spectral function at the interval $[-5, -4]$ is shown in \cref{fig:lambda_1} (top panel). We  see that even such a small deviation from the true value of $\lambda_1$ can result in large errors in analytic continuation. 

The reason behind this is as follows. Let us keep other poles fixed at their true values and let $\lambda_1$ vary from $-6$ to $-4$, and plot the landscape of the loss function $\mathcal E(\{\lambda_l\})$ in \cref{fig:lambda_1} (bottom panel).
We can see that the loss function is highly nonconvex. When $\lambda^{\text{in}}_1$ is outside the interval $[-5.25,-4.5]$, the loss function is essentially flat. In fact, if $\lambda^{\text{in}}_1\notin [-5.25,-4.5]$, the optimized semidefinite matrix $\mathbb X_1^{\mathrm{opt}}$ is a zero matrix. This will result in a vanishing gradient for $\lambda_1^{\text{in}}$ (see \cref{eq:polegradient}), and therefore the optimization could not continue.
 In such an adversarial scenario, a random initialization of poles cannot recover the true pole locations.

\subsection{Bosonic response functions}
\label{subsec:bosonic}
In order to demonstrate that the PES method is also applicable to bosonic response functions, let us consider the quantity $b(\tau) = \langle \hat n_{0\uparrow}(\tau)\hat n_{0\uparrow}(0) \rangle$ of the Hubbard dimer model, which indicates the same-spin susceptibility for the $0$-th site of the system. This function is also  used in \cite{SchottLochtLundin2016} to test the performance of  analytic continuation methods for bosonic functions. 
We use the Hubbard dimer example with the same parameters as in \cref{subsec:hubbarddimer} other than taking $U =0.9$.

Let us use $b(\omega)$ to denote the time-response function in the frequency domain. We are trying to fit the Matsubara data of $b(\tau)$ into the following form:
\begin{equation}
    b(w) = \sum_{l=1}^{N_p}\frac{x_l}{w-\lambda_l},
\end{equation}
where the  quantities $x_l$ satisfies the semidefinite constraints \cref{eq:bosonSDconditions} and the constraint
\begin{equation}
\begin{aligned}
\sum_{l}{x}_{l} & =\frac{1}{Z} \sum_{r, s}\left|\left\langle\Psi_s\left|\hat n_{0\uparrow}\right| \Psi_r\right\rangle\right|^2\left(\mathrm{e}^{-\beta E_s} - \mathrm{e}^{-\beta E_r}\right) = 0.
\end{aligned}
\end{equation} 
Here $N_{\text{orb}}=1$ since we are only considering the time response function with respect to the $0$-th site.

\begin{figure}[htbp]
    \centering
    \includegraphics[width = 85mm]{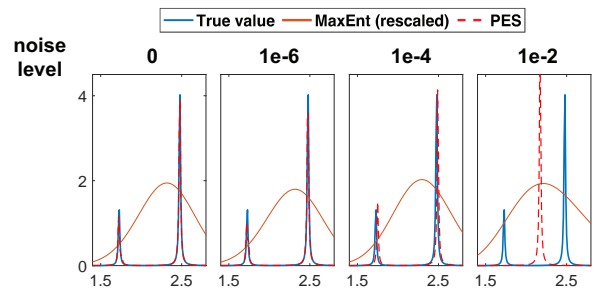}
    \caption{Analytic continuation of bosonic response functions.}
    \label{fig:bosonic}
\end{figure}

In the current physical system, the spectral function of $b(w)$
has four peaks, with two on the positive half axis, and two on the negative half axis. In fact, $b(w)$ is an odd function. Since the  susceptibility  is mirrored for negative frequencies, we only choose to plot the positive half axis. 
Note that 
 since the Nevanlinna and Carath\'{e}odory continuation are not applicable to bosonic functions, we can only compare the PES method to the MaxEnt continuation method, for Matsubara data with different noise levels $\sigma = 0,10^{-6}, 10^{-4}$ and $10^{-2}$. The result is shown in \cref{fig:bosonic}.

The MaxEnt method could not resolve the peaks of the bosonic response function under any noise level.
We can see that when the noise level is small ($\sigma = 0, 10^{-6}, 10^{-4}$), the PES method could recover both peaks on the positive half axis. When the noise level is large ($\sigma = 10^{-2}$), the PES method could only recover one peak. This means that the noise-robustness of PES method for bosonic functions are comparable to that for fermionic functions.

\subsection{Band structure}
\label{subsec:bandstructure}
We now applied the PES method to \textit{ab initio} band structure calculations of solids.
We use two data sets obtained from GW calculations: the matrix-valued Green's functions for crystalline Si in the equilibrium geometry, previously analyzed in Ref.~\cite{FeiYehZgidEtAl2021}, and  for AgI obtained with a relativistic exact two-component formulation in the one-electron approximation (x2c1e) \cite{YehSheeSunGullZgid2022}.
The data have been obtained by fully self-consistent GW formulated in Gaussian Bloch orbitals, as described in detail in Ref.~\cite{YehIskakovZgidGull2022}. 
Both AgI and Si employ  the gth-dzvp-molopt-sr basis \cite{VandeVondeleHutter2007} with gth-pbe pseudopotential \cite{GoedeckerTeterHutter1996}, using integrals generated by the PySCF package \cite{PYSCF} and a code based on ALPS \cite{ALPS}. Calculations are performed on a 6 $\times$ 6 $\times$ 6 grid \cite{IskakovYehGulLZgid2020} and interpolated using a Wannier interpolation of Matsubara data \cite{YehIskakovZgidGull2022} on 52  positive IR frequencies \cite{ShinaokaOtsukiOhzekiYoshimi2017,LiWallerbergerChikanoYehGullShinaoka2020} on a grid with 200 momentum points along a high-symmetry path. The broadening parameter of $\eta = 0.01 \text{Ha} \approx 3157 \text{K}$  is larger than the temperature of $T = 0.001 \text{Ha} \approx 316 \text{K}$ and correlation effects. 

\begin{figure*}[t]
\includegraphics[width=147mm]{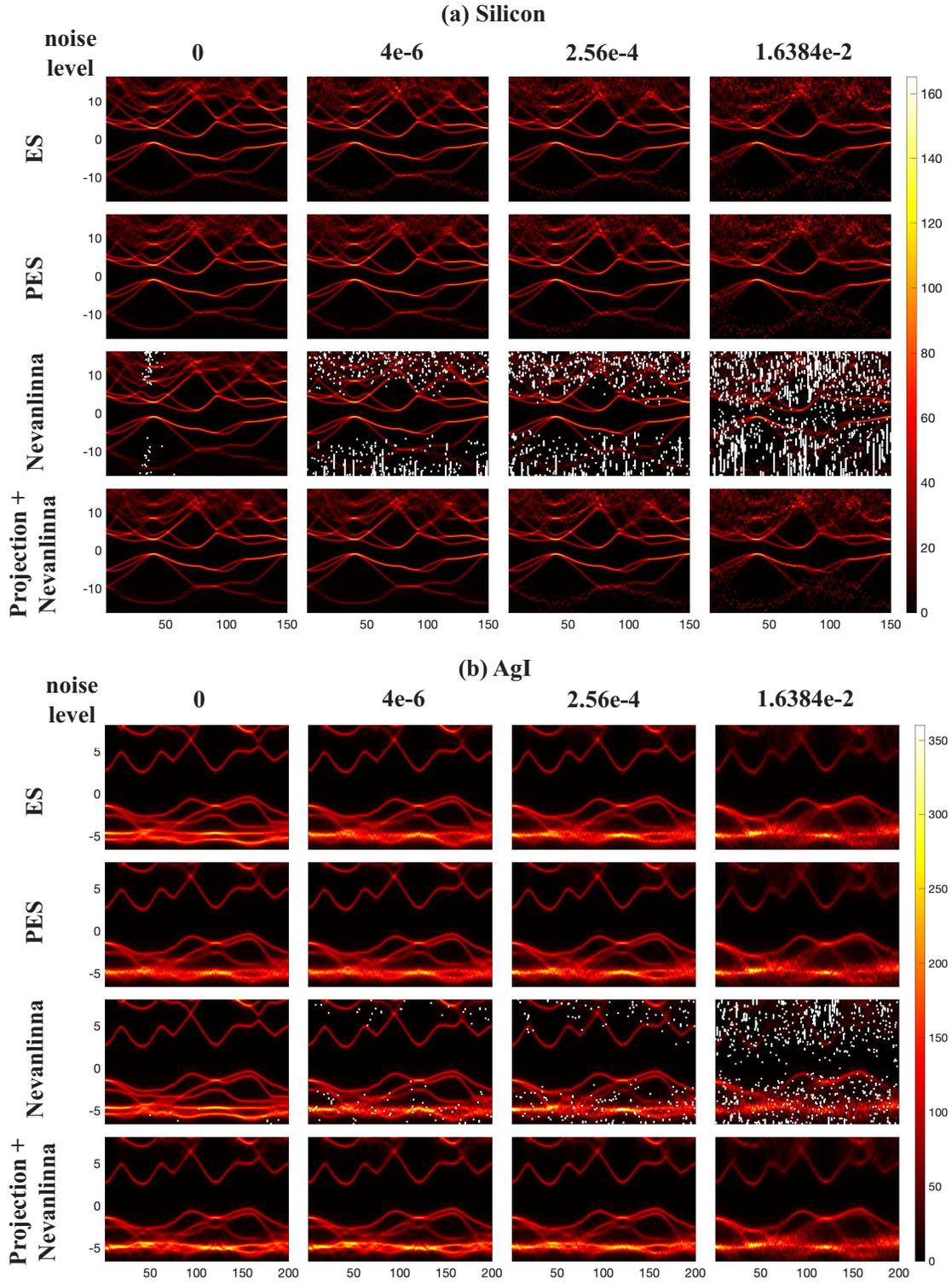}
\caption{Band structure of  Silicon and AgI, revealed by the analytic continuation.  Data from Ref.\cite{YehIskakovZgidGull2022}. White entries correspond to negative (non-causal) spectral functions.}
\label{fig:solid}
\end{figure*}

The analytic continuation is conducted as follows. Since the number of orbitals is  much larger than that in the Hubbard dimer example ($N_{\text{orb}}\approx 70$ for silicon and $N_{\text{orb}}\approx 230$ for AgI), the computational cost will be too high if we conduct the analytic continuation for the entire matrix. Therefore we only implement the analytic continuation for the diagonal elements. The projection step is conducted on each diagonal entry separately with a fine grid of $M=12000$ points.
After the projection step, we estimate the poles of each diagonal entry using AAA algorithm, and then conduct the SDR fitting of each diagonal entry separately.

We conduct analytic continuation for Matsubara data with different  noise levels $\sigma = 0,10^{-6}, 10^{-4}$ and $10^{-2}$.
The results of the spectral functions along a high symmetry path  reveal the band structure, and are plotted in  \cref{fig:solid}.
We place a white marker in the plot if the corresponding component of the spectral function becomes negative to indicate that the result is unphysical.

Once again, we can see that Nevanlinna continuations could result in negative spectral functions even for the clean data, while Nevanlinna continuations combined with the causal projection  can largely resolve the isuse of noise-sensitivity.
 PES method, ES method and the Nevanlinna continuation on projected data routine are all noise-robust. Even at a relatively large noise-level $\sigma = 10^{-2}$, the structure of the valence and conduction bands can be recovered. This means that our noise-robust analytic continuation methods could successfully retrieve information such as the band gap from Matsubara data even under a significant noise level.

\section{Discussion}

In this article, we have proposed a robust analytic continuation method called PES, consisting of the following three steps: causal projection, pole estimation and SDR fitting through bi-level optimization. The PES method is noise-robust, could deal with matrix-valued Green's functions and is applicable to both fermionic and bosonic systems. As a by product, we also find that the causal projection could significantly improve the performance of  noise-sensitive methods such as  Nevanlinna and Carath\'{e}odory continuation.
We have also demonstrated these properties with extensive numerical experiments.

 Our current strategy of pole estimation is to perform AAA algorithm on the projected data. Note that the AAA algorithm is still an interpolation procedure can be susceptible to noise. The causal projection alleviates this drawback, and in practice it often provides a good enough initial guess for SDR fitting even in the noisy scenario. A better strategy to estimate the pole locations (e.g., by combining with the Prony method~\cite{Ying2022pole}) may still be helpful in improving the overall performance of the PES method. Moreover, the current SDP solver (SDPT3) can be computationally expensive when the number of variables is large. In such a case, more efficient solvers based on first-order methods may reduce the computational time in both the projection and the SDR fitting steps. 

As we have shown, the PES method is useful for retrieving the excitation spectra of molecules and band structures in solids. These data feature multiple sharp peaks modeled by poles on the real axis. It could also be used to perform bath fitting in the Dynamical Mean Field Theory (DMFT) \cite{Mejuto-ZaeraZepeda-NunezLindseyEtAl2020}, and it is applicable in lattice gauge theory (for example, see  \cite{BurnierRothkopf2013,TripoltGublerUlybyshevSmekal2019}).
Broadened spectral functions that cannot be resolved into discrete peaks may be modeled by a relatively small number of poles away from the real axis. This would require modification of the definition of the causal space.
Note that in the noisy scenario, it may be very difficult to distinguish the spectra described by a pole in the complex plane and the spectra due to many poles on or near the real axis. 
Therefore prior information such as the maximum number of complex poles and approximate pole locations may become necessary. This scenario will be investigated in future work.

\vspace{1em}
\textbf{Acknowledgment}

This material is based upon work supported by the U.S. Department of Energy, Office of Science, Office of Advanced Scientific Computing Research and Office of Basic Energy Sciences, Scientific Discovery through Advanced Computing (SciDAC) program under Award Number DE-SC0022198 (Z.H.) and DE-SC0022088 (E.G.). The work is partially supported by the Air Force Office of Scientific Research under award number FA9550-18-1-0095 (L.L.). L.L. is a Simons Investigator. We thank Jiahao Yao, Lexing Ying, Yang Yu and Dominika Zgid for helpful discussions. 

\bibliographystyle{apsrev}
\bibliography{main}

\begin{thebibliography}{63}
\expandafter\ifx\csname natexlab\endcsname\relax\def\natexlab#1{#1}\fi
\expandafter\ifx\csname bibnamefont\endcsname\relax
  \def\bibnamefont#1{#1}\fi
\expandafter\ifx\csname bibfnamefont\endcsname\relax
  \def\bibfnamefont#1{#1}\fi
\expandafter\ifx\csname citenamefont\endcsname\relax
  \def\citenamefont#1{#1}\fi
\expandafter\ifx\csname url\endcsname\relax
  \def\url#1{\texttt{#1}}\fi
\expandafter\ifx\csname urlprefix\endcsname\relax\def\urlprefix{URL }\fi
\providecommand{\bibinfo}[2]{#2}
\providecommand{\eprint}[2][]{\url{#2}}

\bibitem[{\citenamefont{Baker et~al.}(1996)\citenamefont{Baker, Baker~Jr,
  Graves-Morris, Baker, and Baker}}]{BakerBakerJrGraves-MorrisEtAl1996}
\bibinfo{author}{\bibfnamefont{G.~A.} \bibnamefont{Baker}},
  \bibinfo{author}{\bibfnamefont{G.~A.} \bibnamefont{Baker~Jr}},
  \bibinfo{author}{\bibfnamefont{P.}~\bibnamefont{Graves-Morris}},
  \bibinfo{author}{\bibfnamefont{G.}~\bibnamefont{Baker}}, \bibnamefont{and}
  \bibinfo{author}{\bibfnamefont{S.~S.} \bibnamefont{Baker}},
  \emph{\bibinfo{title}{{Pade Approximants: Encyclopedia of Mathematics and
  It's Applications, Vol. 59 George A. Baker, Jr., Peter Graves-Morris}}},
  vol.~\bibinfo{volume}{59} (\bibinfo{publisher}{Cambridge University Press},
  \bibinfo{year}{1996}).

\bibitem[{\citenamefont{Sch{\"o}tt et~al.}(2016)\citenamefont{Sch{\"o}tt,
  Locht, Lundin, Gr{\aa}n{\"a}s, Eriksson, and
  Di~Marco}}]{SchottLochtLundin2016}
\bibinfo{author}{\bibfnamefont{J.}~\bibnamefont{Sch{\"o}tt}},
  \bibinfo{author}{\bibfnamefont{I.~L.} \bibnamefont{Locht}},
  \bibinfo{author}{\bibfnamefont{E.}~\bibnamefont{Lundin}},
  \bibinfo{author}{\bibfnamefont{O.}~\bibnamefont{Gr{\aa}n{\"a}s}},
  \bibinfo{author}{\bibfnamefont{O.}~\bibnamefont{Eriksson}}, \bibnamefont{and}
  \bibinfo{author}{\bibfnamefont{I.}~\bibnamefont{Di~Marco}},
  \bibinfo{journal}{Physical Review B} \textbf{\bibinfo{volume}{93}},
  \bibinfo{pages}{075104} (\bibinfo{year}{2016}).

\bibitem[{\citenamefont{Jarrell and Gubernatis}(1996)}]{JarrellGubernatis1996}
\bibinfo{author}{\bibfnamefont{M.}~\bibnamefont{Jarrell}} \bibnamefont{and}
  \bibinfo{author}{\bibfnamefont{J.~E.} \bibnamefont{Gubernatis}},
  \bibinfo{journal}{Physics Reports} \textbf{\bibinfo{volume}{269}},
  \bibinfo{pages}{133} (\bibinfo{year}{1996}).

\bibitem[{\citenamefont{Asakawa et~al.}(2001)\citenamefont{Asakawa, Nakahara,
  and Hatsuda}}]{AsakawaNakaharaHatsuda2001}
\bibinfo{author}{\bibfnamefont{M.}~\bibnamefont{Asakawa}},
  \bibinfo{author}{\bibfnamefont{Y.}~\bibnamefont{Nakahara}}, \bibnamefont{and}
  \bibinfo{author}{\bibfnamefont{T.}~\bibnamefont{Hatsuda}},
  \bibinfo{journal}{Progress in Particle and Nuclear Physics}
  \textbf{\bibinfo{volume}{46}}, \bibinfo{pages}{459} (\bibinfo{year}{2001}).

\bibitem[{\citenamefont{Levy et~al.}(2017)\citenamefont{Levy, LeBlanc, and
  Gull}}]{LevyLeBlancGull2017}
\bibinfo{author}{\bibfnamefont{R.}~\bibnamefont{Levy}},
  \bibinfo{author}{\bibfnamefont{J.}~\bibnamefont{LeBlanc}}, \bibnamefont{and}
  \bibinfo{author}{\bibfnamefont{E.}~\bibnamefont{Gull}},
  \bibinfo{journal}{Computer Physics Communications}
  \textbf{\bibinfo{volume}{215}}, \bibinfo{pages}{149} (\bibinfo{year}{2017}).

\bibitem[{\citenamefont{Bryan}(1990)}]{Bryan1990}
\bibinfo{author}{\bibfnamefont{R.}~\bibnamefont{Bryan}},
  \bibinfo{journal}{European Biophysics Journal} \textbf{\bibinfo{volume}{18}},
  \bibinfo{pages}{165} (\bibinfo{year}{1990}).

\bibitem[{\citenamefont{Rothkopf}(2020)}]{Rothkopf2020}
\bibinfo{author}{\bibfnamefont{A.}~\bibnamefont{Rothkopf}},
  \bibinfo{journal}{Data} \textbf{\bibinfo{volume}{5}}, \bibinfo{pages}{85}
  (\bibinfo{year}{2020}).

\bibitem[{\citenamefont{Sandvik}(1998)}]{Sandvik1998}
\bibinfo{author}{\bibfnamefont{A.~W.} \bibnamefont{Sandvik}},
  \bibinfo{journal}{Physical Review B} \textbf{\bibinfo{volume}{57}},
  \bibinfo{pages}{10287} (\bibinfo{year}{1998}).

\bibitem[{\citenamefont{Vafayi and Gunnarsson}(2007)}]{VafayiGunnarsson2007}
\bibinfo{author}{\bibfnamefont{K.}~\bibnamefont{Vafayi}} \bibnamefont{and}
  \bibinfo{author}{\bibfnamefont{O.}~\bibnamefont{Gunnarsson}},
  \bibinfo{journal}{Physical Review B} \textbf{\bibinfo{volume}{76}},
  \bibinfo{pages}{035115} (\bibinfo{year}{2007}).

\bibitem[{\citenamefont{Fuchs et~al.}(2010)\citenamefont{Fuchs, Pruschke, and
  Jarrell}}]{FuchsPruschkeJarrell2010}
\bibinfo{author}{\bibfnamefont{S.}~\bibnamefont{Fuchs}},
  \bibinfo{author}{\bibfnamefont{T.}~\bibnamefont{Pruschke}}, \bibnamefont{and}
  \bibinfo{author}{\bibfnamefont{M.}~\bibnamefont{Jarrell}},
  \bibinfo{journal}{Physical Review E} \textbf{\bibinfo{volume}{81}},
  \bibinfo{pages}{056701} (\bibinfo{year}{2010}).

\bibitem[{\citenamefont{Shao and Sandvik}(2022)}]{ShaoSandvik2022}
\bibinfo{author}{\bibfnamefont{H.}~\bibnamefont{Shao}} \bibnamefont{and}
  \bibinfo{author}{\bibfnamefont{A.~W.} \bibnamefont{Sandvik}},
  \emph{\bibinfo{title}{Progress on stochastic analytic continuation of quantum
  monte carlo data}} (\bibinfo{year}{2022}), \eprint{2202.09870}.

\bibitem[{\citenamefont{Yoshimi et~al.}(2019)\citenamefont{Yoshimi, Otsuki,
  Motoyama, Ohzeki, and Shinaoka}}]{YoshimiOtsukiMotoyamaEtAl2019}
\bibinfo{author}{\bibfnamefont{K.}~\bibnamefont{Yoshimi}},
  \bibinfo{author}{\bibfnamefont{J.}~\bibnamefont{Otsuki}},
  \bibinfo{author}{\bibfnamefont{Y.}~\bibnamefont{Motoyama}},
  \bibinfo{author}{\bibfnamefont{M.}~\bibnamefont{Ohzeki}}, \bibnamefont{and}
  \bibinfo{author}{\bibfnamefont{H.}~\bibnamefont{Shinaoka}},
  \bibinfo{journal}{Computer Physics Communications}
  \textbf{\bibinfo{volume}{244}}, \bibinfo{pages}{319} (\bibinfo{year}{2019}).

\bibitem[{\citenamefont{Motoyama et~al.}(2022)\citenamefont{Motoyama, Yoshimi,
  and Otsuki}}]{MotoyamaYoshimiOtsuki2022}
\bibinfo{author}{\bibfnamefont{Y.}~\bibnamefont{Motoyama}},
  \bibinfo{author}{\bibfnamefont{K.}~\bibnamefont{Yoshimi}}, \bibnamefont{and}
  \bibinfo{author}{\bibfnamefont{J.}~\bibnamefont{Otsuki}},
  \bibinfo{journal}{Physical Review B} \textbf{\bibinfo{volume}{105}},
  \bibinfo{pages}{035139} (\bibinfo{year}{2022}).

\bibitem[{\citenamefont{Fournier et~al.}(2020)\citenamefont{Fournier, Wang,
  Yazyev, and Wu}}]{FournierWangYazyevEtAl2020}
\bibinfo{author}{\bibfnamefont{R.}~\bibnamefont{Fournier}},
  \bibinfo{author}{\bibfnamefont{L.}~\bibnamefont{Wang}},
  \bibinfo{author}{\bibfnamefont{O.~V.} \bibnamefont{Yazyev}},
  \bibnamefont{and} \bibinfo{author}{\bibfnamefont{Q.}~\bibnamefont{Wu}},
  \bibinfo{journal}{Physical Review Letters} \textbf{\bibinfo{volume}{124}},
  \bibinfo{pages}{056401} (\bibinfo{year}{2020}).

\bibitem[{\citenamefont{Yoon et~al.}(2018)\citenamefont{Yoon, Sim, and
  Han}}]{YoonSimHan2018}
\bibinfo{author}{\bibfnamefont{H.}~\bibnamefont{Yoon}},
  \bibinfo{author}{\bibfnamefont{J.-H.} \bibnamefont{Sim}}, \bibnamefont{and}
  \bibinfo{author}{\bibfnamefont{M.~J.} \bibnamefont{Han}},
  \bibinfo{journal}{Physical Review B} \textbf{\bibinfo{volume}{98}},
  \bibinfo{pages}{245101} (\bibinfo{year}{2018}).

\bibitem[{\citenamefont{Tomczak and Biermann}(2007)}]{TomczakBiermann2007}
\bibinfo{author}{\bibfnamefont{J.~M.} \bibnamefont{Tomczak}} \bibnamefont{and}
  \bibinfo{author}{\bibfnamefont{S.}~\bibnamefont{Biermann}},
  \bibinfo{journal}{Journal of Physics: Condensed Matter}
  \textbf{\bibinfo{volume}{19}}, \bibinfo{pages}{365206}
  (\bibinfo{year}{2007}).

\bibitem[{\citenamefont{Dang et~al.}(2014)\citenamefont{Dang, Ai, Millis, and
  Marianetti}}]{DangAiMillisetal2014}
\bibinfo{author}{\bibfnamefont{H.~T.} \bibnamefont{Dang}},
  \bibinfo{author}{\bibfnamefont{X.}~\bibnamefont{Ai}},
  \bibinfo{author}{\bibfnamefont{A.~J.} \bibnamefont{Millis}},
  \bibnamefont{and} \bibinfo{author}{\bibfnamefont{C.~A.}
  \bibnamefont{Marianetti}}, \bibinfo{journal}{Physical Review B}
  \textbf{\bibinfo{volume}{90}}, \bibinfo{pages}{125114}
  (\bibinfo{year}{2014}).

\bibitem[{\citenamefont{Gull and Millis}(2014)}]{GullMillis2014}
\bibinfo{author}{\bibfnamefont{E.}~\bibnamefont{Gull}} \bibnamefont{and}
  \bibinfo{author}{\bibfnamefont{A.~J.} \bibnamefont{Millis}},
  \bibinfo{journal}{Physical Review B} \textbf{\bibinfo{volume}{90}},
  \bibinfo{pages}{041110} (\bibinfo{year}{2014}).

\bibitem[{\citenamefont{Kraberger et~al.}(2017)\citenamefont{Kraberger, Triebl,
  Zingl, and Aichhorn}}]{krabergerTrieblZingletal2017}
\bibinfo{author}{\bibfnamefont{G.~J.} \bibnamefont{Kraberger}},
  \bibinfo{author}{\bibfnamefont{R.}~\bibnamefont{Triebl}},
  \bibinfo{author}{\bibfnamefont{M.}~\bibnamefont{Zingl}}, \bibnamefont{and}
  \bibinfo{author}{\bibfnamefont{M.}~\bibnamefont{Aichhorn}},
  \bibinfo{journal}{Physical Review B} \textbf{\bibinfo{volume}{96}},
  \bibinfo{pages}{155128} (\bibinfo{year}{2017}).

\bibitem[{\citenamefont{Reymbaut et~al.}(2017)\citenamefont{Reymbaut, Gagnon,
  Bergeron, and Tremblay}}]{ReymbautTremblay2017}
\bibinfo{author}{\bibfnamefont{A.}~\bibnamefont{Reymbaut}},
  \bibinfo{author}{\bibfnamefont{A.-M.} \bibnamefont{Gagnon}},
  \bibinfo{author}{\bibfnamefont{D.}~\bibnamefont{Bergeron}}, \bibnamefont{and}
  \bibinfo{author}{\bibfnamefont{A.-M.~S.} \bibnamefont{Tremblay}},
  \bibinfo{journal}{Phys. Rev. B} \textbf{\bibinfo{volume}{95}},
  \bibinfo{pages}{121104} (\bibinfo{year}{2017}).

\bibitem[{\citenamefont{Kac and Krein}(1974)}]{KacKrein1974}
\bibinfo{author}{\bibfnamefont{I.~S.} \bibnamefont{Kac}} \bibnamefont{and}
  \bibinfo{author}{\bibfnamefont{M.~G.} \bibnamefont{Krein}},
  \bibinfo{journal}{Amer. Math. Soc. Transl} \textbf{\bibinfo{volume}{103}},
  \bibinfo{pages}{1–18} (\bibinfo{year}{1974}).

\bibitem[{\citenamefont{Belyi et~al.}(2006)\citenamefont{Belyi, Hassi, de~Snoo,
  and Tsekanovskii}}]{BelyiHassiSnooEtAl2006}
\bibinfo{author}{\bibfnamefont{S.}~\bibnamefont{Belyi}},
  \bibinfo{author}{\bibfnamefont{S.}~\bibnamefont{Hassi}},
  \bibinfo{author}{\bibfnamefont{H.}~\bibnamefont{de~Snoo}}, \bibnamefont{and}
  \bibinfo{author}{\bibfnamefont{E.}~\bibnamefont{Tsekanovskii}},
  \bibinfo{journal}{Linear algebra and its applications}
  \textbf{\bibinfo{volume}{419}}, \bibinfo{pages}{331} (\bibinfo{year}{2006}).

\bibitem[{\citenamefont{Fei et~al.}(2021{\natexlab{a}})\citenamefont{Fei, Yeh,
  and Gull}}]{FeiYehGull2021}
\bibinfo{author}{\bibfnamefont{J.}~\bibnamefont{Fei}},
  \bibinfo{author}{\bibfnamefont{C.-N.} \bibnamefont{Yeh}}, \bibnamefont{and}
  \bibinfo{author}{\bibfnamefont{E.}~\bibnamefont{Gull}},
  \bibinfo{journal}{Physical Review Letters} \textbf{\bibinfo{volume}{126}},
  \bibinfo{pages}{056402} (\bibinfo{year}{2021}{\natexlab{a}}).

\bibitem[{\citenamefont{Fei et~al.}(2021{\natexlab{b}})\citenamefont{Fei, Yeh,
  Zgid, and Gull}}]{FeiYehZgidEtAl2021}
\bibinfo{author}{\bibfnamefont{J.}~\bibnamefont{Fei}},
  \bibinfo{author}{\bibfnamefont{C.-N.} \bibnamefont{Yeh}},
  \bibinfo{author}{\bibfnamefont{D.}~\bibnamefont{Zgid}}, \bibnamefont{and}
  \bibinfo{author}{\bibfnamefont{E.}~\bibnamefont{Gull}},
  \bibinfo{journal}{Physical Review B} \textbf{\bibinfo{volume}{104}},
  \bibinfo{pages}{165111} (\bibinfo{year}{2021}{\natexlab{b}}).

\bibitem[{\citenamefont{Pick}(1917)}]{Pick1917}
\bibinfo{author}{\bibfnamefont{G.}~\bibnamefont{Pick}},
  \bibinfo{journal}{Mathematische annalen} \textbf{\bibinfo{volume}{78}},
  \bibinfo{pages}{270} (\bibinfo{year}{1917}).

\bibitem[{\citenamefont{Khargonekar and
  Tannenbaum}(1985)}]{KhargonekarTannenbaum1985}
\bibinfo{author}{\bibfnamefont{P.}~\bibnamefont{Khargonekar}} \bibnamefont{and}
  \bibinfo{author}{\bibfnamefont{A.}~\bibnamefont{Tannenbaum}},
  \bibinfo{journal}{IEEE Transactions on Automatic Control}
  \textbf{\bibinfo{volume}{30}}, \bibinfo{pages}{1005} (\bibinfo{year}{1985}).

\bibitem[{\citenamefont{Tannenbaum}(1987)}]{Tannenbaum1987}
\bibinfo{author}{\bibfnamefont{A.}~\bibnamefont{Tannenbaum}}, in
  \emph{\bibinfo{booktitle}{26th IEEE Conference on Decision and Control}}
  (\bibinfo{organization}{IEEE}, \bibinfo{year}{1987}),
  vol.~\bibinfo{volume}{26}, pp. \bibinfo{pages}{1635--1636}.

\bibitem[{\citenamefont{Doyle et~al.}(2013)\citenamefont{Doyle, Francis, and
  Tannenbaum}}]{DoyleFrancisTannenbaum2013}
\bibinfo{author}{\bibfnamefont{J.~C.} \bibnamefont{Doyle}},
  \bibinfo{author}{\bibfnamefont{B.~A.} \bibnamefont{Francis}},
  \bibnamefont{and} \bibinfo{author}{\bibfnamefont{A.~R.}
  \bibnamefont{Tannenbaum}}, \emph{\bibinfo{title}{{Feedback control theory}}}
  (\bibinfo{publisher}{Courier Corporation}, \bibinfo{year}{2013}).

\bibitem[{\citenamefont{Osborne and Smyth}(1995)}]{OsborneSmyth1995}
\bibinfo{author}{\bibfnamefont{M.~R.} \bibnamefont{Osborne}} \bibnamefont{and}
  \bibinfo{author}{\bibfnamefont{G.~K.} \bibnamefont{Smyth}},
  \bibinfo{journal}{SIAM J. Sci. Comput.} \textbf{\bibinfo{volume}{16}},
  \bibinfo{pages}{119} (\bibinfo{year}{1995}).

\bibitem[{\citenamefont{Golub and Pereyra}(2003)}]{GolubPereyra2003}
\bibinfo{author}{\bibfnamefont{G.}~\bibnamefont{Golub}} \bibnamefont{and}
  \bibinfo{author}{\bibfnamefont{V.}~\bibnamefont{Pereyra}},
  \bibinfo{journal}{Inverse problems} \textbf{\bibinfo{volume}{19}},
  \bibinfo{pages}{R1} (\bibinfo{year}{2003}).

\bibitem[{\citenamefont{Beylkin and Monz{\'o}n}(2005)}]{BeylkinMonzon2005}
\bibinfo{author}{\bibfnamefont{G.}~\bibnamefont{Beylkin}} \bibnamefont{and}
  \bibinfo{author}{\bibfnamefont{L.}~\bibnamefont{Monz{\'o}n}},
  \bibinfo{journal}{Applied and Computational Harmonic Analysis}
  \textbf{\bibinfo{volume}{19}}, \bibinfo{pages}{17} (\bibinfo{year}{2005}).

\bibitem[{\citenamefont{Ying}(2022{\natexlab{a}})}]{Ying2022analytic}
\bibinfo{author}{\bibfnamefont{L.}~\bibnamefont{Ying}}, \bibinfo{journal}{arXiv
  preprint arXiv:2202.09719}  (\bibinfo{year}{2022}{\natexlab{a}}).

\bibitem[{\citenamefont{Ying}(2022{\natexlab{b}})}]{Ying2022pole}
\bibinfo{author}{\bibfnamefont{L.}~\bibnamefont{Ying}}, \bibinfo{journal}{arXiv
  preprint arXiv:2202.02670}  (\bibinfo{year}{2022}{\natexlab{b}}).

\bibitem[{\citenamefont{Nakatsukasa et~al.}(2018)\citenamefont{Nakatsukasa,
  S{\`e}te, and Trefethen}}]{NakatsukasaSeteTrefethen2018}
\bibinfo{author}{\bibfnamefont{Y.}~\bibnamefont{Nakatsukasa}},
  \bibinfo{author}{\bibfnamefont{O.}~\bibnamefont{S{\`e}te}}, \bibnamefont{and}
  \bibinfo{author}{\bibfnamefont{L.~N.} \bibnamefont{Trefethen}},
  \bibinfo{journal}{SIAM Journal on Scientific Computing}
  \textbf{\bibinfo{volume}{40}}, \bibinfo{pages}{A1494} (\bibinfo{year}{2018}).

\bibitem[{\citenamefont{Mejuto-Zaera et~al.}(2020)\citenamefont{Mejuto-Zaera,
  Zepeda-N{\'u}{\~n}ez, Lindsey, Tubman, Whaley, and
  Lin}}]{Mejuto-ZaeraZepeda-NunezLindseyEtAl2020}
\bibinfo{author}{\bibfnamefont{C.}~\bibnamefont{Mejuto-Zaera}},
  \bibinfo{author}{\bibfnamefont{L.}~\bibnamefont{Zepeda-N{\'u}{\~n}ez}},
  \bibinfo{author}{\bibfnamefont{M.}~\bibnamefont{Lindsey}},
  \bibinfo{author}{\bibfnamefont{N.}~\bibnamefont{Tubman}},
  \bibinfo{author}{\bibfnamefont{B.}~\bibnamefont{Whaley}}, \bibnamefont{and}
  \bibinfo{author}{\bibfnamefont{L.}~\bibnamefont{Lin}},
  \bibinfo{journal}{Physical Review B} \textbf{\bibinfo{volume}{101}},
  \bibinfo{pages}{035143} (\bibinfo{year}{2020}).

\bibitem[{\citenamefont{Negele and Orland}(1988)}]{NegeleOrland1988}
\bibinfo{author}{\bibfnamefont{J.~W.} \bibnamefont{Negele}} \bibnamefont{and}
  \bibinfo{author}{\bibfnamefont{H.}~\bibnamefont{Orland}},
  \emph{\bibinfo{title}{{Quantum many-particle systems}}}
  (\bibinfo{publisher}{Westview}, \bibinfo{year}{1988}).

\bibitem[{\citenamefont{Gesztesy and
  Tsekanovskii}(2000)}]{GesztesyTsekanovskii2000}
\bibinfo{author}{\bibfnamefont{F.}~\bibnamefont{Gesztesy}} \bibnamefont{and}
  \bibinfo{author}{\bibfnamefont{E.}~\bibnamefont{Tsekanovskii}},
  \bibinfo{journal}{Mathematische Nachrichten} \textbf{\bibinfo{volume}{218}},
  \bibinfo{pages}{61} (\bibinfo{year}{2000}).

\bibitem[{\citenamefont{Shohat and Tamarkin}(1950)}]{ShohatTamarkin1950}
\bibinfo{author}{\bibfnamefont{J.~A.} \bibnamefont{Shohat}} \bibnamefont{and}
  \bibinfo{author}{\bibfnamefont{J.~D.} \bibnamefont{Tamarkin}},
  \emph{\bibinfo{title}{{The problem of moments}}}, vol.~\bibinfo{volume}{1}
  (\bibinfo{publisher}{American Mathematical Society (RI)},
  \bibinfo{year}{1950}).

\bibitem[{\citenamefont{Akhiezer}(2020)}]{Akhiezer2020}
\bibinfo{author}{\bibfnamefont{N.~I.} \bibnamefont{Akhiezer}},
  \emph{\bibinfo{title}{{The classical moment problem and some related
  questions in analysis}}} (\bibinfo{publisher}{SIAM}, \bibinfo{year}{2020}).

\bibitem[{\citenamefont{Grant and Boyd}(2014)}]{GrantBoyd2014}
\bibinfo{author}{\bibfnamefont{M.}~\bibnamefont{Grant}} \bibnamefont{and}
  \bibinfo{author}{\bibfnamefont{S.}~\bibnamefont{Boyd}},
  \emph{\bibinfo{title}{{CVX: Matlab software for disciplined convex
  programming, version 2.1}}} (\bibinfo{year}{2014}).

\bibitem[{\citenamefont{Tibshirani}(1996)}]{Tibshirani1996}
\bibinfo{author}{\bibfnamefont{R.}~\bibnamefont{Tibshirani}},
  \bibinfo{journal}{J. Royal Stat. Soc. B} \textbf{\bibinfo{volume}{58}},
  \bibinfo{pages}{267} (\bibinfo{year}{1996}).

\bibitem[{\citenamefont{Candes et~al.}(2006)\citenamefont{Candes, Romberg, and
  Tao}}]{CandesRombergTao2006}
\bibinfo{author}{\bibfnamefont{E.~J.} \bibnamefont{Candes}},
  \bibinfo{author}{\bibfnamefont{J.~K.} \bibnamefont{Romberg}},
  \bibnamefont{and} \bibinfo{author}{\bibfnamefont{T.}~\bibnamefont{Tao}},
  \bibinfo{journal}{Communications on Pure and Applied Mathematics}
  \textbf{\bibinfo{volume}{59}}, \bibinfo{pages}{1207} (\bibinfo{year}{2006}).

\bibitem[{\citenamefont{{Bach, Francis and Jenatton, Rodolph and Mairal, Julien
  and Obozinski, Guillaume}}(2012)}]{BachJenattonMairal2012}
\bibinfo{author}{\bibnamefont{{Bach, Francis and Jenatton, Rodolph and Mairal,
  Julien and Obozinski, Guillaume}}} (\bibinfo{year}{2012}).

\bibitem[{\citenamefont{Driscoll et~al.}(2014)\citenamefont{Driscoll, Hale, and
  Trefethen}}]{DriscollHaleTrefethen2014}
\bibinfo{author}{\bibfnamefont{T.~A.} \bibnamefont{Driscoll}},
  \bibinfo{author}{\bibfnamefont{N.}~\bibnamefont{Hale}}, \bibnamefont{and}
  \bibinfo{author}{\bibfnamefont{L.~N.} \bibnamefont{Trefethen}},
  \emph{\bibinfo{title}{{Chebfun guide}}} (\bibinfo{year}{2014}).

\bibitem[{\citenamefont{Berrut and Trefethen}(2004)}]{BerrutTrefethen2004}
\bibinfo{author}{\bibfnamefont{J.-P.} \bibnamefont{Berrut}} \bibnamefont{and}
  \bibinfo{author}{\bibfnamefont{L.~N.} \bibnamefont{Trefethen}},
  \bibinfo{journal}{SIAM Rev.} \textbf{\bibinfo{volume}{46}},
  \bibinfo{pages}{501} (\bibinfo{year}{2004}).

\bibitem[{\citenamefont{Feynman}(1939)}]{Feynman1939}
\bibinfo{author}{\bibfnamefont{R.~P.} \bibnamefont{Feynman}},
  \bibinfo{journal}{Physical review} \textbf{\bibinfo{volume}{56}},
  \bibinfo{pages}{340} (\bibinfo{year}{1939}).

\bibitem[{\citenamefont{Nocedal and Wright}(1999)}]{NocedalWright1999}
\bibinfo{author}{\bibfnamefont{J.}~\bibnamefont{Nocedal}} \bibnamefont{and}
  \bibinfo{author}{\bibfnamefont{S.~J.} \bibnamefont{Wright}},
  \emph{\bibinfo{title}{Numerical optimization}} (\bibinfo{publisher}{Springer
  Verlag}, \bibinfo{year}{1999}).

\bibitem[{\citenamefont{Toh et~al.}(2012)\citenamefont{Toh, Todd, and
  T{\"u}t{\"u}nc{\"u}}}]{TohToddTutuncu2012}
\bibinfo{author}{\bibfnamefont{K.-C.} \bibnamefont{Toh}},
  \bibinfo{author}{\bibfnamefont{M.~J.} \bibnamefont{Todd}}, \bibnamefont{and}
  \bibinfo{author}{\bibfnamefont{R.~H.} \bibnamefont{T{\"u}t{\"u}nc{\"u}}}, in
  \emph{\bibinfo{booktitle}{Handbook on semidefinite, conic and polynomial
  optimization}} (\bibinfo{publisher}{Springer}, \bibinfo{year}{2012}), pp.
  \bibinfo{pages}{715--754}.

\bibitem[{\citenamefont{Mehrotra}(1992)}]{Mehrotra1992}
\bibinfo{author}{\bibfnamefont{S.}~\bibnamefont{Mehrotra}},
  \bibinfo{journal}{SIAM Journal on optimization} \textbf{\bibinfo{volume}{2}},
  \bibinfo{pages}{575} (\bibinfo{year}{1992}).

\bibitem[{\citenamefont{Vandenberghe and Boyd}(1996)}]{VandenbergheBoyd1996}
\bibinfo{author}{\bibfnamefont{L.}~\bibnamefont{Vandenberghe}}
  \bibnamefont{and} \bibinfo{author}{\bibfnamefont{S.}~\bibnamefont{Boyd}},
  \bibinfo{journal}{SIAM Rev.} \textbf{\bibinfo{volume}{38}},
  \bibinfo{pages}{49} (\bibinfo{year}{1996}).

\bibitem[{\citenamefont{{Jiang, Kaifeng and Sun, Defeng and Toh,
  Kim-Chuan}}(2012)}]{JiangSunToh2012}
\bibinfo{author}{\bibnamefont{{Jiang, Kaifeng and Sun, Defeng and Toh,
  Kim-Chuan}}}, \bibinfo{journal}{SIAM J. Optim.}
  \textbf{\bibinfo{volume}{22}}, \bibinfo{pages}{1042} (\bibinfo{year}{2012}).

\bibitem[{\citenamefont{Nakatsukasa and
  Trefethen}(2020)}]{NakatsukasaTrefethen2020}
\bibinfo{author}{\bibfnamefont{Y.}~\bibnamefont{Nakatsukasa}} \bibnamefont{and}
  \bibinfo{author}{\bibfnamefont{L.~N.} \bibnamefont{Trefethen}},
  \bibinfo{journal}{SIAM Journal on Scientific Computing}
  \textbf{\bibinfo{volume}{42}}, \bibinfo{pages}{A3157} (\bibinfo{year}{2020}).

\bibitem[{\citenamefont{Yeh et~al.}(2022{\natexlab{a}})\citenamefont{Yeh, Shee,
  Sun, Gull, and Zgid}}]{YehSheeSunGullZgid2022}
\bibinfo{author}{\bibfnamefont{C.-N.} \bibnamefont{Yeh}},
  \bibinfo{author}{\bibfnamefont{A.}~\bibnamefont{Shee}},
  \bibinfo{author}{\bibfnamefont{Q.}~\bibnamefont{Sun}},
  \bibinfo{author}{\bibfnamefont{E.}~\bibnamefont{Gull}}, \bibnamefont{and}
  \bibinfo{author}{\bibfnamefont{D.}~\bibnamefont{Zgid}}
  (\bibinfo{year}{2022}{\natexlab{a}}).

\bibitem[{\citenamefont{Yeh et~al.}(2022{\natexlab{b}})\citenamefont{Yeh,
  Iskakov, Zgid, and Gull}}]{YehIskakovZgidGull2022}
\bibinfo{author}{\bibfnamefont{C.-N.} \bibnamefont{Yeh}},
  \bibinfo{author}{\bibfnamefont{S.}~\bibnamefont{Iskakov}},
  \bibinfo{author}{\bibfnamefont{D.}~\bibnamefont{Zgid}}, \bibnamefont{and}
  \bibinfo{author}{\bibfnamefont{E.}~\bibnamefont{Gull}},
  \emph{\bibinfo{title}{Fully self-consistent finite-temperature $gw$ in
  gaussian bloch orbitals for solids}} (\bibinfo{year}{2022}{\natexlab{b}}).

\bibitem[{\citenamefont{VandeVondele and
  Hutter}(2007)}]{VandeVondeleHutter2007}
\bibinfo{author}{\bibfnamefont{J.}~\bibnamefont{VandeVondele}}
  \bibnamefont{and} \bibinfo{author}{\bibfnamefont{J.}~\bibnamefont{Hutter}},
  \bibinfo{journal}{The Journal of Chemical Physics}
  \textbf{\bibinfo{volume}{127}}, \bibinfo{pages}{114105}
  (\bibinfo{year}{2007}), \eprint{https://doi.org/10.1063/1.2770708}.

\bibitem[{\citenamefont{Goedecker et~al.}(1996)\citenamefont{Goedecker, Teter,
  and Hutter}}]{GoedeckerTeterHutter1996}
\bibinfo{author}{\bibfnamefont{S.}~\bibnamefont{Goedecker}},
  \bibinfo{author}{\bibfnamefont{M.}~\bibnamefont{Teter}}, \bibnamefont{and}
  \bibinfo{author}{\bibfnamefont{J.}~\bibnamefont{Hutter}},
  \bibinfo{journal}{Phys. Rev. B} \textbf{\bibinfo{volume}{54}},
  \bibinfo{pages}{1703} (\bibinfo{year}{1996}).

\bibitem[{\citenamefont{Sun et~al.}(2020)\citenamefont{Sun, Zhang, Banerjee,
  Bao, Barbry, Blunt, Bogdanov, Booth, Chen, Cui et~al.}}]{PYSCF}
\bibinfo{author}{\bibfnamefont{Q.}~\bibnamefont{Sun}},
  \bibinfo{author}{\bibfnamefont{X.}~\bibnamefont{Zhang}},
  \bibinfo{author}{\bibfnamefont{S.}~\bibnamefont{Banerjee}},
  \bibinfo{author}{\bibfnamefont{P.}~\bibnamefont{Bao}},
  \bibinfo{author}{\bibfnamefont{M.}~\bibnamefont{Barbry}},
  \bibinfo{author}{\bibfnamefont{N.~S.} \bibnamefont{Blunt}},
  \bibinfo{author}{\bibfnamefont{N.~A.} \bibnamefont{Bogdanov}},
  \bibinfo{author}{\bibfnamefont{G.~H.} \bibnamefont{Booth}},
  \bibinfo{author}{\bibfnamefont{J.}~\bibnamefont{Chen}},
  \bibinfo{author}{\bibfnamefont{Z.-H.} \bibnamefont{Cui}},
  \bibnamefont{et~al.}, \bibinfo{journal}{The Journal of Chemical Physics}
  \textbf{\bibinfo{volume}{153}}, \bibinfo{pages}{024109}
  (\bibinfo{year}{2020}), \eprint{https://doi.org/10.1063/5.0006074}.

\bibitem[{\citenamefont{Gaenko et~al.}(2017)\citenamefont{Gaenko, Antipov,
  Carcassi, Chen, Chen, Dong, Gamper, Gukelberger, Igarashi, Iskakov
  et~al.}}]{ALPS}
\bibinfo{author}{\bibfnamefont{A.}~\bibnamefont{Gaenko}},
  \bibinfo{author}{\bibfnamefont{A.}~\bibnamefont{Antipov}},
  \bibinfo{author}{\bibfnamefont{G.}~\bibnamefont{Carcassi}},
  \bibinfo{author}{\bibfnamefont{T.}~\bibnamefont{Chen}},
  \bibinfo{author}{\bibfnamefont{X.}~\bibnamefont{Chen}},
  \bibinfo{author}{\bibfnamefont{Q.}~\bibnamefont{Dong}},
  \bibinfo{author}{\bibfnamefont{L.}~\bibnamefont{Gamper}},
  \bibinfo{author}{\bibfnamefont{J.}~\bibnamefont{Gukelberger}},
  \bibinfo{author}{\bibfnamefont{R.}~\bibnamefont{Igarashi}},
  \bibinfo{author}{\bibfnamefont{S.}~\bibnamefont{Iskakov}},
  \bibnamefont{et~al.}, \bibinfo{journal}{Computer Physics Communications}
  \textbf{\bibinfo{volume}{213}}, \bibinfo{pages}{235} (\bibinfo{year}{2017}),
  ISSN \bibinfo{issn}{0010-4655}.

\bibitem[{\citenamefont{Iskakov et~al.}(2020)\citenamefont{Iskakov, Yeh, Gull,
  and Zgid}}]{IskakovYehGulLZgid2020}
\bibinfo{author}{\bibfnamefont{S.}~\bibnamefont{Iskakov}},
  \bibinfo{author}{\bibfnamefont{C.-N.} \bibnamefont{Yeh}},
  \bibinfo{author}{\bibfnamefont{E.}~\bibnamefont{Gull}}, \bibnamefont{and}
  \bibinfo{author}{\bibfnamefont{D.}~\bibnamefont{Zgid}},
  \bibinfo{journal}{Phys. Rev. B} \textbf{\bibinfo{volume}{102}},
  \bibinfo{pages}{085105} (\bibinfo{year}{2020}).

\bibitem[{\citenamefont{Shinaoka et~al.}(2017)\citenamefont{Shinaoka, Otsuki,
  Ohzeki, and Yoshimi}}]{ShinaokaOtsukiOhzekiYoshimi2017}
\bibinfo{author}{\bibfnamefont{H.}~\bibnamefont{Shinaoka}},
  \bibinfo{author}{\bibfnamefont{J.}~\bibnamefont{Otsuki}},
  \bibinfo{author}{\bibfnamefont{M.}~\bibnamefont{Ohzeki}}, \bibnamefont{and}
  \bibinfo{author}{\bibfnamefont{K.}~\bibnamefont{Yoshimi}},
  \bibinfo{journal}{Phys. Rev. B} \textbf{\bibinfo{volume}{96}},
  \bibinfo{pages}{035147} (\bibinfo{year}{2017}).

\bibitem[{\citenamefont{Li et~al.}(2020)\citenamefont{Li, Wallerberger,
  Chikano, Yeh, Gull, and Shinaoka}}]{LiWallerbergerChikanoYehGullShinaoka2020}
\bibinfo{author}{\bibfnamefont{J.}~\bibnamefont{Li}},
  \bibinfo{author}{\bibfnamefont{M.}~\bibnamefont{Wallerberger}},
  \bibinfo{author}{\bibfnamefont{N.}~\bibnamefont{Chikano}},
  \bibinfo{author}{\bibfnamefont{C.-N.} \bibnamefont{Yeh}},
  \bibinfo{author}{\bibfnamefont{E.}~\bibnamefont{Gull}}, \bibnamefont{and}
  \bibinfo{author}{\bibfnamefont{H.}~\bibnamefont{Shinaoka}},
  \bibinfo{journal}{Phys. Rev. B} \textbf{\bibinfo{volume}{101}},
  \bibinfo{pages}{035144} (\bibinfo{year}{2020}).

\bibitem[{\citenamefont{Burnier and Rothkopf}(2013)}]{BurnierRothkopf2013}
\bibinfo{author}{\bibfnamefont{Y.}~\bibnamefont{Burnier}} \bibnamefont{and}
  \bibinfo{author}{\bibfnamefont{A.}~\bibnamefont{Rothkopf}},
  \bibinfo{journal}{Phys. Rev. Lett.} \textbf{\bibinfo{volume}{111}},
  \bibinfo{pages}{182003} (\bibinfo{year}{2013}).

\bibitem[{\citenamefont{Tripolt et~al.}(2019)\citenamefont{Tripolt, Gubler,
  Ulybyshev, and {von Smekal}}}]{TripoltGublerUlybyshevSmekal2019}
\bibinfo{author}{\bibfnamefont{R.-A.} \bibnamefont{Tripolt}},
  \bibinfo{author}{\bibfnamefont{P.}~\bibnamefont{Gubler}},
  \bibinfo{author}{\bibfnamefont{M.}~\bibnamefont{Ulybyshev}},
  \bibnamefont{and} \bibinfo{author}{\bibfnamefont{L.}~\bibnamefont{{von
  Smekal}}}, \bibinfo{journal}{Computer Physics Communications}
  \textbf{\bibinfo{volume}{237}}, \bibinfo{pages}{129} (\bibinfo{year}{2019}),
  ISSN \bibinfo{issn}{0010-4655}.

\end{thebibliography}



\end{document}